\documentclass[11pt,cite]{article}

\usepackage{jheppub}

\usepackage[latin1]{inputenc}

\newcommand{\lbl}[1]{\label{eq:#1}}
\newcommand{ \rf}[1]{(\ref{eq:#1})}

\newcommand{\be}{\begin{equation}}
\newcommand{\ee}{\end{equation}}
\newcommand{\bea}{\begin{eqnarray}}
\newcommand{\eea}{\end{eqnarray}}
\newcommand{\setl}{\setlength\arraycolsep{2pt}}

\newcommand{\noi}{\noindent}
\newcommand{\nn}{\nonumber}
\newcommand{\ra}{\rightarrow}
\newcommand{\Ra}{\Rightarrow}

\newcommand{\cA}{{\cal A}}

\newcommand{\cF}{{\cal F}}

\newcommand{\cO}{{\cal O}}

\newcommand{\cR}{{\cal R}}

\newcommand{\Imm}{\mbox{\rm Im}}
\newcommand{\Ree}{\mbox{\rm Re}}

\newcommand{\tr}{\mbox{\rm tr}}
\newcommand{\Li}{\mbox{\rm Li}}
\newcommand{\MeV}{\mbox{\rm MeV}}
\newcommand{\GeV}{\mbox{\rm GeV}}

\newcommand{\with}{\mbox{\rm with}}

\newcommand{\annd}{\mbox{\rm and}}
\newcommand{\foor}{\mbox{\rm for}}

\newcommand{\als}{\alpha_{\mbox{\rm {\scriptsize s}}}}

\newcommand{\GF}{G_{\mbox{\rm {\tiny F}}}}

\newcommand{\msb}{\overline{\mbox{\rm\footnotesize MS}}}

\newcommand{\elm}{\mbox{\rm {\tiny em}}}
\newcommand{\nc}{\mbox{\rm {\tiny nc}}}
\newcommand{\Nc}{\mbox{${\rm N_c}$}}

\newcommand{\ksls}{\not \! k}

\newcommand{\qsls}{\not \! q}
\newcommand{\pslsh}{\not \! p}

\newcommand{\dd}{\!\cdot\!}

\title{Hadronic Contributions to the Muon Anomaly in the Constituent Chiral Quark Model}

\author[a]{David Greynat}
\author[b,1]{and Eduardo de Rafael.}

\note{Partially based on the talk of EdeR at the INT Workshop on ``The hadronic Light--by--Light Contribution to the Muon Anomaly'', February 28th -March 4th, 2011.}

\affiliation[a]{\it Departamento de F\'isica Te\'orica\\ Facultad de Ciencias, Universidad de Zaragoza\\ E-50009 Zaragoza, Spain} 

\affiliation[b]{\it Centre  de Physique Th{\'e}orique~\footnote{Unit{\'e} Mixte de Recherche (UMR7332) du CNRS et des Universit{\'e}s Aix Marseille 1, Aix Marseille 2 et Sud Toulon-Var, affili{\'e}e {\`a} la FRUMAM.}\\
       CNRS-Luminy, Case 907\\
    F-13288 Marseille Cedex 9, France}

\abstract{The hadronic contributions to the anomalous magnetic moment of the muon which are relevant for the confrontation between theory and experiment at the present level of accuracy,  are  evaluated within the same framework: the constituent chiral quark model. This includes the contributions from the dominant hadronic vacuum polarization as well as  from the next--to--leading order hadronic vacuum polarization, the contributions from the hadronic light-by-light scattering, and the contributions from the electroweak hadronic $Z\gamma\gamma$ vertex. They are all evaluated as a function of only one free parameter: the constituent quark mass. We also comment on the comparison between our results and other phenomenological evaluations.}

\begin{document}

\maketitle

\section{\normalsize Introduction}\lbl{int}
\setcounter{equation}{0}
\def\theequation{\arabic{section}.\arabic{equation}}

The present experimental world average of the anomalous magnetic moment of the muon $a_{\mu}$, assuming CPT--invariance, viz. $a_{\mu^+} = a_{\mu^-}$,  is
\be\lbl{amuexp}
a_{\mu}^{(\rm exp)}=116~592~080~(63)\times 10^{-11}\quad (0.54~{\rm ppm})\,,
\ee
where the total uncertainty includes a 0.46 ppm statistical uncertainty and a 0.28 ppm systematic uncertainty, combined in quadrature.
This result is largely dominated by the latest series of precise measurements carried out  at the Brookhaven National Laboratory (BNL)  by the E821 collaboration, with results reported in ref.~\cite{Bennet06} and references therein. The prediction of the Standard Model, as a result of contributions from many physicists is~\footnote{For recent reviews see e.g.  refs.~\cite{MdeRR07,JN09} and references therein.}
\be\lbl{amuth}
a_{\mu}^{(\rm SM)}=116~591~801~(49)\times 10^{-11}\,,
\ee
where the error here is dominated at present  by the lowest order hadronic vacuum polarization  contribution uncertainty~\cite{DHMZ10} ($\pm 42.0\times 10^{-11}$), as well as by the  contribution from the hadronic light--by--light scattering, which is theoretically estimated to be $(105\pm 26)\times 10^{-11}$~\cite{PdeRV09}.  The results quoted in~\rf{amuexp} and \rf{amuth} imply a significant 3.6 standard deviation between theory and experiment
which deserves attention. In order to firmly attribute this discrepancy to new Physics,  one would like to reduce the theoretical uncertainties as much as possible, parallel to the new experimental efforts towards an even more precise measurement of $a_{\mu}$ in the near future~\cite{Fermi,Jpark}. It is therefore important to reexamine critically  the various theoretical contributions to Eq.~\rf{amuth}; in particular the hadronic contributions. Ideally, one would like to do that within the framework of Quantum Chromodynamics(QCD). Unfortunatley, this demands mastering QCD at all scales from short to long distances, something which is not under full analytic control at present. Therefore, one has to resort to experimental information whenever possible, to QCD inspired hadronic models, and to lattice QCD simulations which are as yet at an early stage. As a result, all the theoretical evaluations of the hadronic contributions to $a_{\mu}$ have systematic errors which are not easy to pin down rigorously. 

Our purpose here is to establish a simple {\it  reference model} to evaluate the various hadronic contributions to $a_{\mu}$ within the same framework, and use it as a yardstick to compare with the more detailed evaluations in the literature.   
The {\it reference model} which we propose is based on  
the Constituent Chiral Quark Model (C$\chi$QM)~\cite{MG84}. This model emerged as an attempt to reconcile  the  successes of phenomenological quark models, like the De~R\'ujula-Georgi-Glashow model~\cite{deRGG75}, with QCD. The corresponding Lagrangian proposed by Manohar and Georgi (MG) is an effective field theory which incorporates the interactions of the low--lying pseudoscalar particles of the hadronic spectrum, the Nambu-Goldstone modes of the spontaneously broken chiral symmetry (S$\chi$SB), to lowest order in the chiral expansion~\cite{Wei79} and in the presence of chirally rotated quark fields.  Because of the S$\chi$SB, these quark fields appear to be massive.
This model, in the presence of $SU(3)_{L}\times SU(3)_{R}$ external sources has been reconsidered recently by one of us~\cite{EdeR11}. As emphasized by Weinberg~\cite{Wei10}, the corresponding effective Lagrangian is renormalizable  in the Large--${\rm N_c}$ limit; however, the number of the required counterterms depends crucially on the value of the coupling constant $g_A$ in the model and, as shown in~\cite{EdeR11}, it is minimized for $g_A =1$. With this choice, and a  value for the constituent quark mass fixed phenomenologically,  
the model reproduces rather well the values of several well known low energy constants.

As discussed in ref.~\cite{EdeR11} the C$\chi$QM model has, however,   its own limitations. Applications to the evaluation of low--energy observables involving the integration of Green's functions over a full range of euclidean momenta fail, in general, because there is no matching of the model to the QCD short--distance behaviour. There is, however,  an exceptional class of  low--energy observables for which  the MG--Lagrangian predictions can be rather reliable. This is the case when the 
leading short--distance behaviour of the underlying Green's function of a given observable is governed by perturbative QCD. The decay $\pi^0 \ra e^+ e^-$, which was discussed in ref.~\cite{EdeR11}, is one such  example.
Other interesting examples of this class of observables   are the contributions to $a_{\mu}$ from Hadronic Vacuum Polarization, from the Hadronic Light--by--Light Scattering and from the Hadronic $Z\gamma\gamma$ vertex ( provided, as we shall see, that the coupling $g_A$ is fixed to $g_A=1$).
The evaluation of these contributions with the C$\chi$QM Lagrangian is the main purpose of this paper and they are discussed below in detail. They have the advantage of simplicity and can provide a consistency check with the more elaborated phenomenological approaches.

\vspace*{1cm}
\section{\normalsize Hadronic Vacuum Polarization.}
\setcounter{equation}{0}
\def\theequation{\arabic{section}.\arabic{equation}}
\noi
There is a well known representation~\cite{BM62} of the dominant contribution to the muon anomaly
from the hadronic vacuum polarization shown in Fig.~\rf{fig:HVP}

\begin{figure}[h]
\begin{center}
\includegraphics[width=0.4\textwidth]{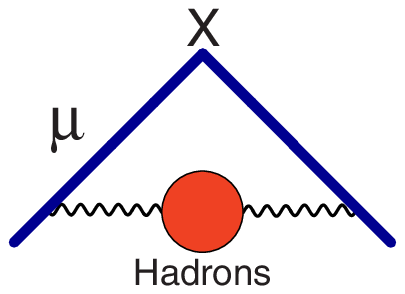}
\bf\caption{\lbl{fig:HVP}}
\vspace*{0.25cm}
{\it  Hadronic Vacuum Polarization contribution to the Muon Anomaly.}
\end{center}
\end{figure}

\be\lbl{GdeR}
\frac{1}{2}(g_{\mu}-2)_{\mbox{\rm\tiny HVP}}\equiv a_{\mu}^{(\rm HVP)}=\int_{4M_{Q}^2}^{\infty}
\frac{dt}{t}K\left(\frac{t}{m_{\mu}^2}\right)
\frac{1}{\pi}\Imm\Pi^{(\rm HVP)}(t)\,,
\ee
where~\footnote{The analytic expression of $K\left(t/m_{\mu}^2\right)$ was first given in ref.~\cite{BdeR68}; see also ref.~\cite{LdeR68}.} 
\be\lbl{BdeR}
K\left(\frac{t}{m_{\mu}^2}\right)=\left(\frac{\alpha}{\pi}\right)\int_{0}^{1}dx\frac{x^2(1-x)}{x^2+\frac{t}{m_{\mu}^2}(1-x)}\,,
\ee
and
$\frac{1}{\pi}\Imm\Pi(t)$ denotes the electromagnetic
hadronic spectral function. It is a useful representation because of the direct relation to the \underline{\it one--photon} $e^+ e^-$ annihilation cross--section into hadrons $(m_e \ra 0)$: 
\begin{equation}\lbl{onephoton}
        \sigma(t)_{\{e^+ e^- \ra (\gamma)\ra {\rm hadrons}\}}=\frac{4\pi^2\alpha}{t} \frac{1}{\pi}
        \Imm\Pi^{({\rm HVP})}(t)\,,
\end{equation}
and hence to experimental data, provided the necessary radiative corrections have been made to insure that one is using the \underline{\it one--photon} cross--section.

In the C$\chi$QM with active $u$, $d$, $s$ quarks and, to a first approximation, with neglect of gluonic corrections

{\setl
\bea\lbl{CQM}
\frac{1}{\pi}\Imm\Pi^{({\rm HVP})}(t) & = & 	\left(\frac{\alpha}{\pi}\right)N_c \left[\left(\frac{2}{3}\right)^2+\left(-\frac{1}{3}\right)^2 +\left(-\frac{1}{3}\right)^2\right] \nn \\
 & \times & \delta \left(\frac{1}{2}-\frac{1}{6}\delta^2\right)\theta(t-4M_Q^2)\,,\quad\with\quad
	\delta=\sqrt{1-\frac{4M_Q^2}{t}}\,.
\eea}

\noi
The integral in Eq.~\rf{GdeR} can then be easily done with the result shown in Fig.~\rf{fig:HVPtotal}, the curve labeled (a), where the value for $a_{\mu}^{(\rm HVP)}$ is plotted as a function of the only free parameter in the model, the constituent quark mass $M_Q$~\footnote{There are many estimates of this contribution, as well as some of the higher order ones, with quark models which can be found in the literature. An earlier reference is \cite{CNPdeR77} and two more recent ones \cite{Pi03} and \cite{ES06}.}

The constituent quark fields in the C$\chi$QM  are assumed to have gluonic interactions as well but, since the Goldstone modes are already in the Lagrangian, the color--${\rm SU(3)}$ coupling constant  is supposed to be no longer running below a scale $\mu_0 \simeq 2~\GeV$ where $\alpha_s (\mu_0)\simeq 0.33$ and non-perturbative effects become significant.  With inclusion of the leading gluonic corrections in perturbation theory, and to leading order in Large--${\rm N_c}$, the spectral function in Eq.~\rf{CQM} then becomes

{\setl
\bea\lbl{CQMg}
\hspace*{-1cm}
\frac{1}{\pi}\Imm\Pi^{({\rm HVP})}(t) & = & 	\left(\frac{\alpha}{\pi}\right)N_c\frac{2}{3}\left\{\delta \left(\frac{1}{2}-\frac{1}{6}\delta^2\right)\right.\nn \\
& + & \left. \left[\frac{N_c \alpha_{\rm s}(\mu_0)}{\pi}\frac{3}{8}\ \theta(\mu_0^2 -t)+\frac{N_c \alpha_{\rm s}(\sqrt{t})}{\pi}\frac{3}{8}\ \theta(t-\mu_0^2)\right]\rho_{\rm KS}(t)\right\}\,,   
\eea}

\noi
where $\rho_{\rm KS}(t)$ can be extracted from the early QED calculation of Källen and Sabry~\cite{KS55} (see also ref.~\cite{LdeR68}):

{\setl
\bea\lbl{KS}
\rho_{\rm KS}(t) & = & 
\delta\left(\frac{5}{8}-\frac{3}{8}\delta^2 -\left(\frac{1}{2}-\frac{1}{6} \delta^2\right)\log\left[64\frac{\delta^4}{(1-\delta^2)^3} \right]\right)\nn \\
  & + & \left(\frac{11}{16}+\frac{11}{24}\delta^2 -\frac{7}{48}\delta^4
+\left(\frac{1}{2}+\frac{1}{3}\delta^2 -\frac{1}{6}\delta^4  \right)\log\left[\frac{(1+\delta)^3}{8\delta^2}\right]\right)\log\left[\frac{1+\delta}{1-\delta} \right] \nn  \\
& + &  2\left(\frac{1}{2}+\frac{1}{3}\delta^2 -\frac{1}{6}\delta^4  \right)
\left(2\ \Li_2\left[\frac{1-\delta}{1+\delta} \right]+\Li_2 \left[-\frac{1-\delta}{1+\delta} \right] \right)\,.
\eea}

\noi
Also, at the level of the accuracy expected from the C$\chi$QM, it is sufficient to  use the  one loop expression
\be
\frac{ \alpha_{\rm s}(\sqrt{t})}{\pi}\simeq\frac{1}{\left(\frac{11}{6}N_c -\frac{n_f}{3}\right)\log\frac{\sqrt{t}}{\Lambda}}\,,\quad{\rm with}\quad \Lambda\simeq 250~\MeV\quad\annd\quad n_f =3\,.
\ee
The resulting value for $a_{\mu}^{(\rm HVP)}$ in Eq.~\rf{GdeR}  in $10^{-10}$ units   versus $M_Q$ in $\MeV$ is shown in Fig.~\rf{fig:HVPtotal}, the curved labeled (b).

In order to compare the C$\chi$QM results for $a_{\mu}^{(\rm HVP)}$ with the phenomenological determinations which incorporate experimental data, we still have to correct for the fact that the curve (b) in Fig.~\rf{fig:HVPtotal} only reflects the Large--{\Nc} estimate of the model. As an estimate of the $1/\Nc$--suppressed effects, we then consider the contributions from the $\pi^+ \pi^-$ and $K^+ K^-$ intermediate states to the spectral function in Eq.~\rf{GdeR}, as predicted by the C$\chi$QM. 
Notice that in this evaluation, the point like coupling
$(-ie) (p^{\mu}-p'^{\mu})$ of scalar QED is replaced by the dressed coupling:
\be
(-ie) (p^{\mu}-p'^{\mu})\Ra (-ie) (p^{\mu}-p'^{\mu})\left\{1+\cF (Q^2)\right\}\,,
\ee
with $\cF (Q^2)$ the pion (kaon) electromagnetic form factor of the C$\chi$QM, at the one loop level in Fig.~\rf{fig:Gpipi}
\begin{figure}[h]
\begin{center}
\includegraphics[width=0.75\textwidth]{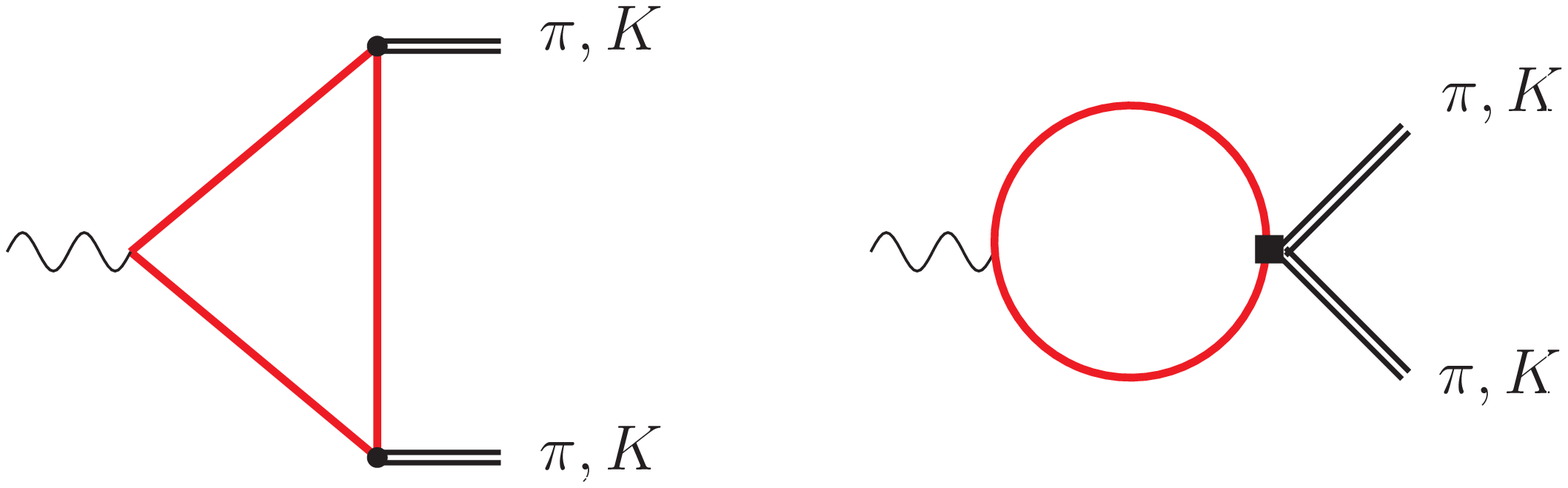}
\bf\caption{\lbl{fig:Gpipi}}
\vspace*{0.25cm}
{\it  Feynman diagrams contributing to the electromagnetic form factor $\cF(Q^2)$ in Eq.~\rf{FFpi}.
}
\end{center}
\end{figure}

\noi
which, for $g_A=1$, is given by the expression:

{\setl
\bea\lbl{FFpi}
\cF (Q^2) & = &  \Nc\frac{Q^2}{16\pi^2 f_{\pi}^2}\int_0^1 dx\ x\int_0^1 dy\ [1-x(1-y)]\frac{M_Q^2}{M_Q^2 +xy(1-x)Q^2 -i\epsilon}\nn \\
 & = &  \frac{\Nc}{16\pi^2}\frac{Q^2}{f_{\pi}^2} \left(-\frac{4M_Q^2}{Q^2}\right)\left[
 1+\frac{1}{2}\sqrt{1+\frac{4 M_Q^2}{Q^2}}\log\frac{\sqrt{1+\frac{4 M_Q^2}{Q^2}}-1}{\sqrt{1+\frac{4 M_Q^2}{Q^2}}+1} \right]\lbl{formf}\,.
\eea}

\noi
In fact, this form factor, for $g_A\not= 1$, has UV--contributions which diverge and would require an explicit counterterm in the Lagrangian. The form factor in Eq.~\rf{formf} has the asymptotic behaviour:
\be
\lim_{ Q^2 \ra 0} \cF (Q^2)= \frac{\Nc}{16\pi^2 f_{\pi}^2}Q^2 \left[ \frac{1}{3}+\frac{1}{30}\frac{Q^2}{M_Q^2}+\cO\left(\frac{Q^4}{M_Q^4} \right)\right]\,,
\ee
and, in particular, fixes the value of the coupling constant $L_9$ in the $\chi$PT  effective Lagrangian to~\cite{EdeRT90}:
\be
L_9 =\frac{\Nc}{16\pi^2 }\frac{1}{3}\,.
\ee 
Also, for $Q^2 =-t$ and $t\ge 4M_Q^2$ the form factor develops an imaginary part:
\be
\frac{1}{\pi}\Imm\cF (t)=  \frac{\Nc}{16\pi^2}\left(-\frac{2M_Q^2}{f_{\pi}^2}\right)
\sqrt{1-\frac{4M_Q^2}{t}}\ \theta(t-4M_Q^2)\,,
\ee
and the form factor $\cF(Q^2)$ obeys a once subtracted dispersion relation
\be
\cF(Q^2)=\int_{4 M_Q^2}^{\infty}\left(\frac{1}{t+Q^2 -i\epsilon}-\frac{1}{t}\right)\frac{1}{\pi}\Imm\cF (t)\,,
\ee
the subtraction ensuring that
$
\cF (Q^2 =0)=0\,,
$
as fixed by lowest order $\chi$PT. 

The  form factor $\cF (Q^2)$, however, does not match the QCD behaviour at large--$Q^2$ values and, therefore, the estimate we propose for the $1/\Nc$--suppresed contributions to the the muon anomaly can only be considered as reasonable up to values of $t$ in Eq.~\rf{GdeR} below $t\sim\mu_0$ where the asymptotic pQCD regime sets in. Contributions beyond $t\sim\mu_0$ have already been taken into account by the second term of the spectral function in Eq.~\rf{CQMg}. 

The total contribution to $a_{\mu}^{(\rm HVP)}$ in the C$\chi$QM, which incorporates gluonic contributions in the spectral function  in Eq.~\rf{CQMg} as well as the subleading  $\pi^+ \pi^-$ and $K^+ K^-$  contributions in the way  described above is shown in Fig.~\rf{fig:HVPtotal} as a function of $M_Q$, the curve labeled (c).

\begin{figure}[h]
\begin{center}
\includegraphics[width=0.75\textwidth]{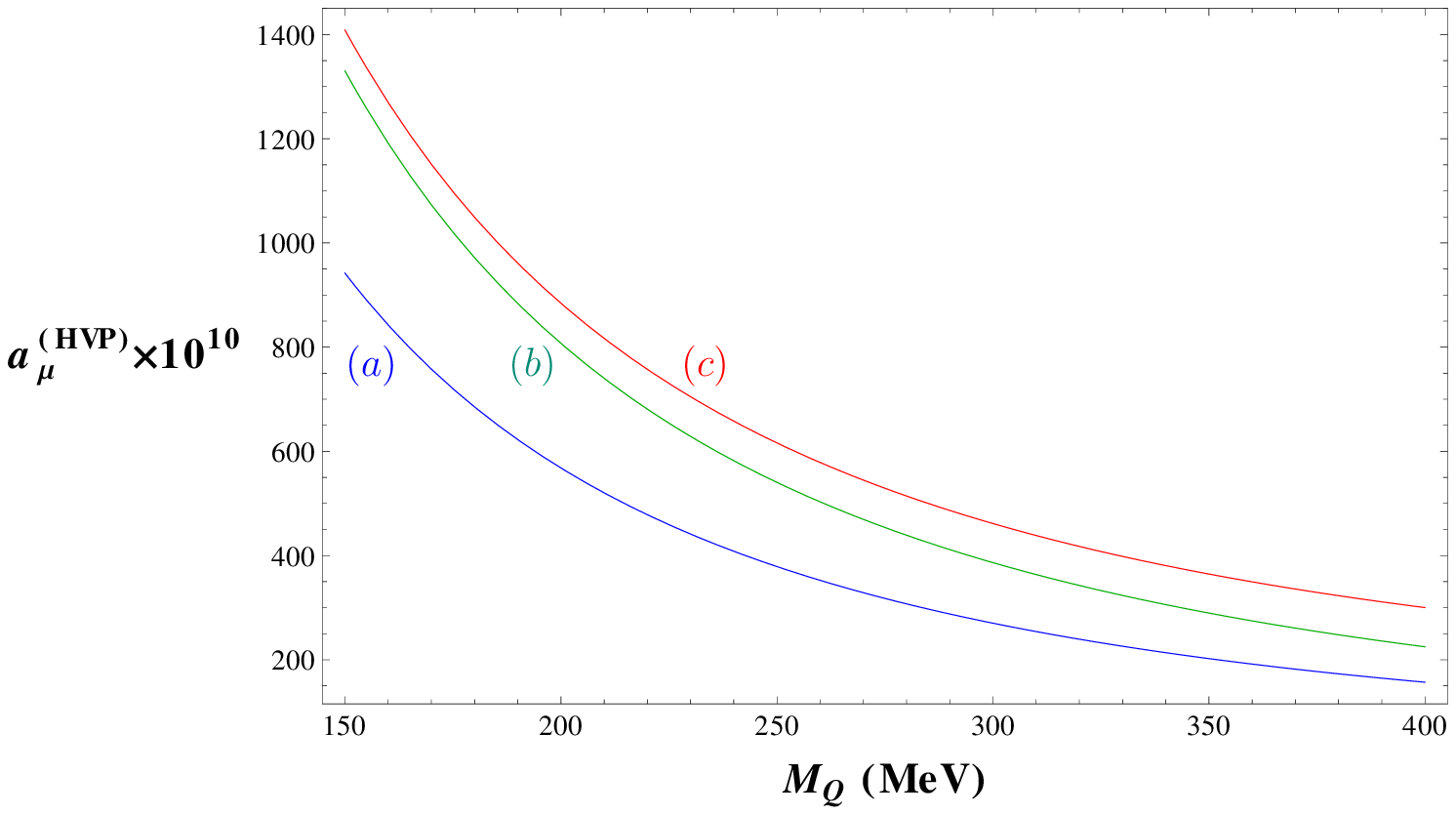}
\bf\caption{\lbl{fig:HVPtotal}}
\end{center}
\vspace*{0.25cm}
{\it\small  $a_{\mu}^{(\rm HVP)}$ in the C$\chi$QM. Curve (a) is the contribution using the spectral function in Eq.~\rf{CQM}; curve (b) the contribution using the corrected spectral function in Eq.~\rf{CQMg} and  curve (c) the contribution using the corrected spectral function in Eq.~\rf{CQMg} 
with subleading  $\pi^+ \pi^-$ and $K^+ K^-$  contributions incorporated as discussed in the text.
}
\end{figure}
 
These considerations provide us with a framework to fix the constituent quark mass $M_Q$. The prediction of the C$\chi$QM, as described above,  should be compared to the phenomenological contribution from hadrons formed of $u$, $d$ and $s$ quarks only, at the level of  one--photon exchange. Contributions like for example the one from an intermediate $\pi^0 \gamma$ state should therefore be excluded so far (more on that later on), as well as those involving $c$, $b$ and $t$ quarks. From the numbers quoted in TABLE II of ref.~\cite{DHMZ10}, we then find that this restriction reduces the phenomenological determination of the anomaly from hadronic vacuum polarization
 to a central value
\be\lbl{exp}
a_{\mu}^{({\rm HVP})}\vert_{\rm phen.}\simeq 653\times 10^{-10}
\ee
which, when compared with the results plotted in Fig.~\rf{fig:HVPtotal}, shows that fixing $M_Q$ in the range 
\be\lbl{MQ}
M_Q =(240\pm 10)~\MeV\,,
\ee
reproduces the phenomenological determination within an error of less than $10\%$.
This determination of the constituent quark mass is the value which we shall  systematically use for $M_Q$  when evaluating the predictions for the other hadronic contributions to the muon anomaly.  We shall then compare them to the various  phenomenological determinations in the literature. 

We wish to emphasize, however, that the error of $10~\MeV$ in Eq.~\rf{MQ} only reflects the phenomenological  choice that we have made in order to fix $M_Q$. As discussed in the Introduction the C$\chi$QM is only  a model of low energy QCD and, as such,  there is no a priori way to fix $M_Q$ from first principles. The error in Eq.~\rf{MQ} does not reflect the systematic error due to other plausible ways of fixing $M_Q$.  

At this stage we wish to point out that the recent lattice QCD determination of $a_{\mu}^{({\rm HVP})}$ with two flavours reported in ref.~\cite{Latt11} can also be very well digested with a value of $M_Q$ within the range given in Eq.~\rf{MQ}.

\vspace*{1cm}
\section{\normalsize Hadronic Vacuum Polarization Contributions at Next--to--Leading Order.}
\setcounter{equation}{0}
\def\theequation{\arabic{section}.\arabic{equation}}
\noi
The Hadronic Vacuum Polarization contributions at $\cO \left(\frac{\alpha}{\pi} \right)^3$ were classified long time ago in ref.~\cite{CNPdeR77}. Let us discuss their evaluation in the C$\chi$QM.

\subsection {\normalsize Class A: 
	{\sc HVP insertions in the fourth order QED vertex diagrams.}}
	
		They correspond to the Feynman diagrams shown in Fig.~\rf{fig:H4}, where the diagrams in  each line in this figure are well-defined gauge invariant subsets. Here, the equivalent of the function $K\left(t/m_{\mu}^2\right)$ in Eq.~\rf{BdeR} at the two loop level, which we call $K^{(4)}\left(t/m_{\mu}^2\right)$, 
was	calculated analytically  by Barbieri and Remiddi~\cite{BR75}. The exact  expression is, however, rather cumbersome and for our purposes it is more convenient to use an expansion of this function in powers of $\frac{m_{\mu}^2}{t}$, which is justified by the fact that the hadronic threshold in the integral that gives the contribution from the diagrams of  Class A: 
\be\lbl{A}
a_{\mu}^{(\rm HVP-A)}=\int_{4M_{Q}^2}^{\infty}
\frac{dt}{t}K^{(4)}(t/m_{\mu}^2)\ 
\frac{1}{\pi}\Imm\Pi^{(\rm HVP)}(t)\,,
\ee
starts at $4M_{Q}^2 \gg m_{\mu}^2$. The terms in the expansion in question for the kernel $K^{(4)}\left(t/m_{\mu}^2\right)$ which we have retained are:

{\setl
\bea
\hspace*{-0.5cm}
K^{(4)}\left(t/m_{\mu}^2\right) & = & \left(\frac{\alpha}{\pi}\right)^2 \left(-2 \frac{m_{\mu}^2}{t}\right)\left\{\left(\frac{23}{36}\log\frac{t}{m_{\mu}^2}  -\frac{223}{54}+\frac{\pi^2}{3}\right)\right.\nn \\
 & &  +\frac{m_{\mu}^2}{t}\left(-\frac{19}{144}\log^2 \frac{t}{m_{\mu}^2}+\frac{367}{216}\log\frac{t}{m_{\mu}^2}-\frac{8785}{1152}+\frac{37}{48}\pi^2  \right)\nn \\
  & &  +\left(\frac{m_{\mu}^2}{t}\right)^2 \left(-\frac{141}{80}\log^2 \frac{t}{m_{\mu}^2}+\frac{10079}{3600}\log \frac{t}{m_{\mu}^2} -\frac{13072841}{432000}+\frac{883}{240}\pi^2\right)\nn \\
  & &  +\left(\frac{m_{\mu}^2}{t}\right)^3 \left(-\frac{961}{80}\log^2 \frac{t}{m_{\mu}^2}  +\frac{6517}{1800}\log \frac{t}{m_{\mu}^2} -\frac{2034703}{16000}+\frac{1301}{80}\pi^2\right) \nn \\
  & & +\left. \cO \left[ \left(\frac{m_{\mu}^2}{t}\right)^4 \right]\right\}\,.
\eea}

\begin{figure}[h]
\begin{center}
\includegraphics[width=0.75\textwidth]{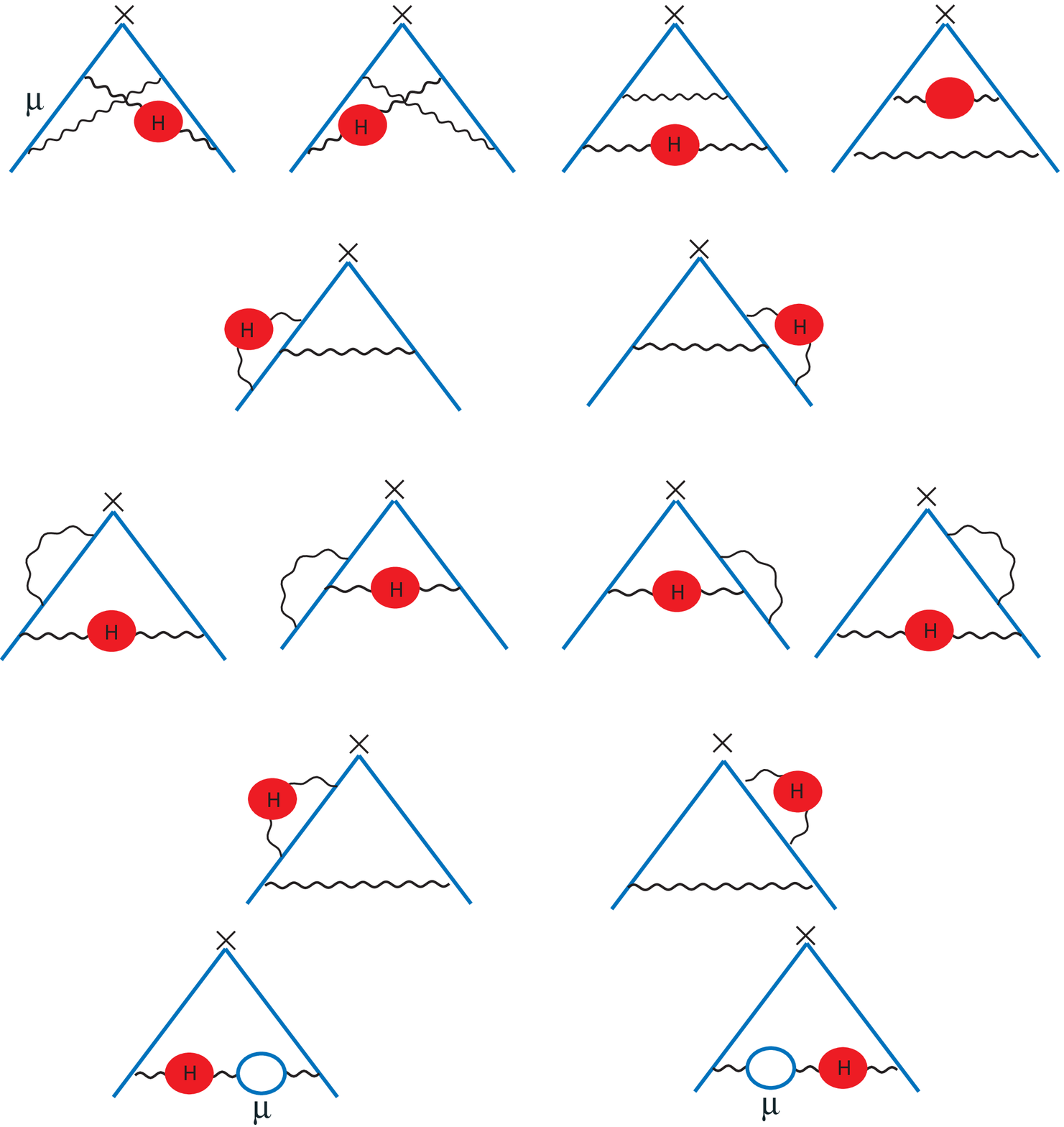}
\bf\caption{\lbl{fig:H4}}
\vspace*{0.25cm}
{\it\small  Feynman diagrams corresponding to the Class A contribution to the muon anomaly.
}
\end{center}
\end{figure}

\noi
Using the C$\chi$QM spectral function in Eq.~\rf{CQMg}, we find for this contribution the following result:
\be
a_{\mu}^{(\rm HVP-A)}(M_Q =240~\MeV)=-171\times 10^{-11}\,,
\ee
with a range 
\be
-181\times 10^{-11} \le a_{\mu}^{(\rm HVP-A)}\le -161\times 10^{-11}\,,\quad \foor\quad 230~\MeV\le M_Q \le250~\MeV\,.
\ee

This result is to be compared with the  phenomenological determination~\cite{Hag11}:
\be
a_{\mu}^{(\rm HVP-A)}=-(207.3\pm 1.9)\times 10^{-11}\,.
\ee
We conclude that the C$\chi$QM reproduces, within the expected accuracy of the model, this phenomenological value, specially if we take into account that the phenomenological determination includes contributions subleading in $1/\Nc$ and from higher flavours, which are beyond the duality domain of the model. 

\subsection {\normalsize Class B:
	 {\sc HVP  insertions in the  QED vertex  with an electron loop.}}
	 
	  This is the contribution from the two Feynman diagrams in Fig.~\rf{fig:eH}. A convenient representation~\cite{CNPdeR77} for this contribution, is the one given by the integral
	  
\begin{figure}[h]
\begin{center}
\includegraphics[width=0.4\textwidth]{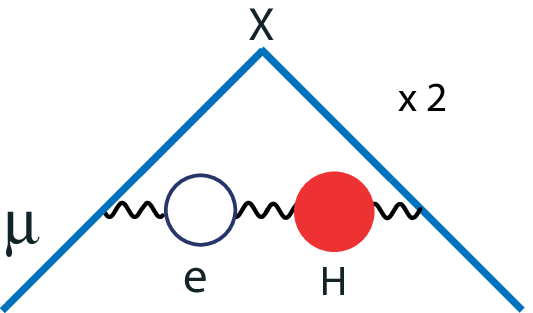}
\bf\caption{\lbl{fig:eH}}
\vspace*{0.25cm}
{\it Feynman diagrams corresponding to the Class B contribution to the muon anomaly.
}
\end{center}
\end{figure}

{\setl
\bea\lbl{B}
a_{\mu}^{(\rm HVP-B)} & = & \left(\frac{\alpha}{\pi}\right)^2  \int_{4M_{Q}^2}^{\infty}
\frac{dt}{t}\int_{0}^{1}dx\frac{x^2(1-x)}{x^2+\frac{t}{m_{\mu}^2}(1-x)}\nn \\
 & \times &  
\left[-2\ \Ree\Pi^{(e)}\left(-\frac{x^2}{1-x}m_{\mu}^2 \right)\frac{1}{\pi}\Imm\Pi^{(\rm HVP)}(t)\right]\,,
\eea}

\noi
where
\be
\Ree\Pi^{(e)}\left(-\frac{x^2}{1-x}m_{\mu}^2 \right)=\frac{8}{9}-\frac{1}{3}\beta^2
+\beta\left(\frac{1}{2}-\frac{1}{6}\beta^2 \right)\log\left( \frac{-1+\beta}{1+\beta}\right)
\ee
with 
\be
\beta=\sqrt{1+4\frac{1-x}{x^2}\frac{m_e^2}{m_{\mu}^2}}\,,
\ee
denotes the real part of the electron self--energy.
Using the C$\chi$QM spectral function in Eq.~\rf{CQMg}, we find for this contribution the following results:
\be
a_{\mu}^{(\rm HVP-B)}(M_Q =240~\MeV)=88.9\times 10^{-11}\,,
\ee
with a range 
\be
82.6\times 10^{-11} \le a_{\mu}^{(\rm HVP-B)}\le 95.9\times 10^{-11}\,,\quad \foor\quad 250~\MeV\ge M_Q \ge 230~\MeV\,.
\ee

This result is to be compared with the phenomenological determination of this contribution which gives~\cite{Hag11}:
\be
a_{\mu}^{(\rm HVP-B)}=(106.0\pm 0.9)\times 10^{-11}\,.
\ee
We find again that, within the expected accuracy of the model, the C$\chi$QM reproduces the phenomenological determination.

\subsection {\normalsize Class C:
{\sc Iterated HVP Contributions.}}

\begin{figure}[h]
\begin{center}
\includegraphics[width=0.4\textwidth]{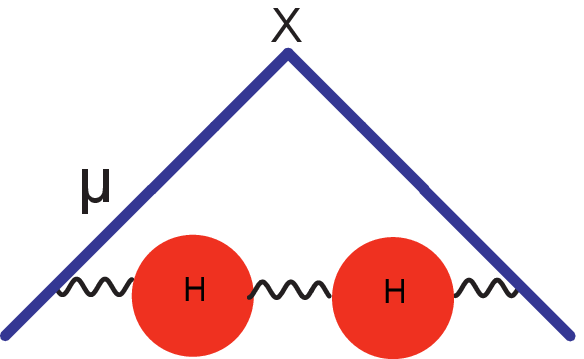}
\bf\caption{\lbl{fig:H2}}
\vspace*{0.25cm}
{\it  HVP Contributions at $\cO(\alpha)$.
}
\end{center}
\end{figure}

This is the contribution in Fig.~\rf{fig:H2}  induced by the quadratic term in the expansion of the photon propagator in the lowest order vertex, fully dressed by hadronic vacuum polarization corrections, i.e. 
\be\lbl{fullpp}
D_{\alpha\beta}^{({\rm HVP})}(q)=-i\left(g_{\alpha\beta}-\frac{q_{\alpha}q_{\beta}}{q^2}\right)\frac{1}{q^2} \frac{1}{1+\Pi^{({\rm HVP})}(q^2)}-ia\frac{q_{\alpha}q_{\beta}}{q^4}\,,
\ee
where $\Pi^{({\rm HVP})}(q^2)$ denotes the proper vacuum polarization self--energy  contribution induced by hadrons and $a$ is a parameter reflecting the gauge 
freedom in the free--field propagator ($a=1$ in the Feynman gauge). In fact, since the diagrams we are considering are gauge independent,  terms proportional to $q_{\alpha}q_{\beta}$ do not contribute to their evaluation.
The lowest order muon anomaly is then modified as follows:
\be
\left(\frac{\alpha}{\pi}\right)\frac{1}{2}\Ra  \left(\frac{\alpha}{\pi}\right)\int_0^1 dx (1-x) \frac{1}{1+\Pi^{({\rm HVP})}\left(\frac{-x^2}{1-x}m_{\mu}^2\right)}\,,
 \ee
and the perturbation theory expansion generates a  series in powers of the self--energy function  $\Pi^{({\rm HVP})}\left(\frac{-x^2}{1-x}m_{\mu}^2\right)$:
\be\lbl{vpex}
 \frac{1}{1+\Pi^{({\rm HVP})}\left(\frac{-x^2}{1-x}m_{\mu}^2\right)} = 1 -\Pi^{({\rm HVP})}\left(\frac{-x^2}{1-x}m_{\mu}^2\right) + \left[\Pi^{({\rm HVP})}\left(\frac{-x^2}{1-x}m_{\mu}^2\right)\right]^2 +\cdots\,.
\ee 
Writing a dispersion relation for {\it each power} of  $\Pi^{({\rm HVP})}\left(\frac{-x^2}{1-x} m_{\mu}^2\right)$ to lowest order in the electromagnetic hadronic interaction, i.e.,
\be
\Pi^{({\rm HVP})}\left(\frac{-x^2}{1-x} m_{\mu}^2\right)= \int_{4m_{\pi}^2}
^{\infty} \frac{dt}{t}\ \frac{\frac{-x^2}{1-x} m_{\mu}^2}{t - \frac{-x^2}{1-x} m_{\mu}^2-i\epsilon}\ \frac{1}{\pi}\Imm\Pi^{({\rm HVP})}\left(t\right)\,,
\ee
results then, from the quadratic term of the expansion in Eq.~\rf{vpex}, in the following representation~\cite{CNPdeR77} for the contribution from the Feynman diagrams in Fig.~6
\be\lbl{rep}
 a_{\mu}^{({\rm HVP-C})}= 	 \int_{4m_{\pi}^2}^{\infty}\frac{dt}{t}\frac{1}{\pi}\Imm\Pi^{({\rm HVP})}\left(t\right) \int_{4m_{\pi}^2}^{\infty}\frac{dt'}{t'}\frac{1}{\pi}\Imm\Pi^{({\rm HVP})}\left(t'\right) 
K\left(\frac{t}{m_{\mu}^2},\frac{t'}{m_{\mu}^2}\right)\,,
\ee
with 
\be\lbl{kernel}
K\left(\frac{t}{m_{\mu}^2},\frac{t'}{m_{\mu}^2}\right)= \left(\frac{\alpha}{\pi}\right)\int_0^1 dx\frac{x^4 (1-x)}{\left(x^2 +\frac{t}{m_{\mu}^2}(1-x)\right)\left(x^2 +\frac{t'}{m_{\mu}^2}(1-x)\right)}\,,
\ee
a composite kernel which correlates the two spectral functions. 
Using the C$\chi$QM spectral function in Eq.~\rf{CQMg} for both $\frac{1}{\pi}\Imm\Pi^{({\rm HVP})}\left(t\right)$ 
and $\frac{1}{\pi}\Imm\Pi^{({\rm HVP})}\left(t'\right)$, we find a small contribution from this C--class:
\be
a_{\mu}^{(\rm HVP-C)}(M_Q =240~\MeV)=2.2\times 10^{-11}\,,
\ee
with a range 
\be
1.9\times 10^{-11} \le a_{\mu}^{(\rm HVP-C)}\le 2.5\times 10^{-11}\,,\quad \foor\quad 250~\MeV\ge M_Q \ge 230~\MeV\,.
\ee

Two independent phenomenological determinations of this contribution (with errors which are likely to have been underestimated) give:
\be
a_{\mu}^{(\rm HVP-C)}({\rm ref.}~\text{\cite{Hag11}})=(3.4\pm 0.1)\times 10^{-11}\,,
\ee
and
\be
a_{\mu}^{(\rm HVP-C)}({\rm ref.}~\text{\cite{Jeger06}})=(3.0\pm 0.1)\times 10^{-11}\,.
\ee
Again, they compare reasonably well with the C$\chi$QM prediction.

\begin{itemize}
\item{\sc Why is this contribution so small?} 

This is an interesting question which, to our knowledge, has not been addressed in the literature. We wish to take the opportunity to answer it here.

The main point is the following: instead of writing a dispersion relation for   
{\it each power} of  $\Pi^{{\rm (HVP)}}\left(q^2\right)$, we could have chosen to write a dispersion relation for the {\it squared photon self--energy}  $\left[\Pi^{(HVP)}(q^2)\right]^2$, i.e. 
\be\lbl{disp2}
	\left[\Pi^{{\rm (HVP)}}(q^2)\right]^2 =\int_0^\infty \frac{dt}{t}\frac{q^2}{t-q^2-i\epsilon}\ 2\  \Ree\Pi^{{\rm (HVP)}}(t)\frac{1}{\pi}\ \Imm\Pi^{{\rm (HVP)}}(t)\,.
\ee
This leads to a representation for the muon anomaly ($q^2 \equiv \frac{-x^2}{1-x} m_{\mu}^2$):
\be\lbl{rep2}\hspace*{-0.5cm}
a_{\mu}^{({\rm HVP-C})}= \left(\frac{\alpha}{\pi}\right)\!\!\!   \int_{4m_{\pi}^2}^{\infty}\!\frac{dt}{t}\int_0^1 \! dx \frac{x^2 (1-x)}{x^2 +\frac{t}{m_{\mu}^2} (1-x)}
\left[-2\  \Ree\Pi^{{\rm (HVP)}}(t)\ \frac{1}{\pi}\Imm\Pi^{{\rm (HVP)}}(t)\right]\,,
\ee
similar to the one we have used in Eq.~\rf{B} for the evaluation of 
$a_{\mu}^{(\rm HVP-B)}$.
Gauge invariance guarantees that the subtraction constant in the double dispersion relation

{\setl
\bea\lbl{dispdisp}
\lefteqn{\Pi^{{\rm (HVP)}}(q^2)\times \Pi^{{\rm (HVP)}}(q^2)=}\nn \\ & & 	\int_0^\infty \frac{dt}{t}\frac{q^2}{t-q^2-i\epsilon}\ \frac{1}{\pi}\Imm\Pi^{{\rm (HVP)}}(t)\times 	\int_0^\infty \frac{dt'}{t'}\frac{q^2}{t'-q^2-i\epsilon}\ \frac{1}{\pi}\Imm\Pi^{{\rm (HVP)}}(t')\,
\eea}

\noi
and the one in the single dispersion relation in Eq.~\rf{disp2} are the same, so that the physical electric charge corresponds to the   one measured classically. In other words, gauge invariance guarantees that the physical content of the two equations \rf{dispdisp} and \rf{disp2} must be the same. Yet, algebraically, starting with the r.h.s. in Eq.~\rf{dispdisp}
 and using the partial fraction decomposition: 
\begin{equation}	\frac{1}{t-q^2-i\epsilon}\ \frac{1}{t'-q^2 -i\epsilon}=\frac{1}{t-q^2-i\epsilon}\ \frac{1}{t'-t}+\frac{1}{t'-q^2-i\epsilon}\ \frac{1}{t-t'}\,,
\end{equation}
one gets the following relation:
\be
\hspace*{-0.2cm}	\Pi^{{\rm (HVP)}}(q^2)\times \Pi^{{\rm (HVP)}}(q^2)\!  = \!
\left[\Pi^{{\rm (HVP)}}(q^2)\right]^2	   
\!\!\! -q^2\!\! \int_0^\infty\! \frac{dt}{t^2}\ 2 \Ree\Pi^{{\rm (HVP)}}(t)\frac{1}{\pi}\Imm\Pi^{{\rm (HVP)}}(t)\,.
\ee
Obviously, the only way to preserve the identity $\Pi^{{\rm (HVP)}}(q^2)\times \Pi^{{\rm (HVP)}}(q^2)  = 
\left[\Pi^{{\rm (HVP)}}(q^2)\right]^2$  is that
\begin{equation}\lbl{dyncon}
	\int_0^\infty\frac{dt}{t^2}\ 2\ \Ree\Pi^{{\rm (HVP)}}(t)\frac{1}{\pi}\Imm\Pi^{{\rm (HVP)}}(t)=0\,,
\end{equation}
which is a highly non trivial constraint!~\footnote{In QED, one can easily check this constraint in perturbation theory.}.
It is this constraint which answers the question of why  $a_{\mu}^{({\rm HVP-C})}$ turns out to be so small. 
Indeed, it implies that the {\it a priori} leading term of $\cO(m_{\mu}^2)$ in an expansion in powers of $m_{\mu}^2$ in the r.h.s. of Eq.~\rf{rep2}, contrary to what happens with the lowest order hadronic vacuum polarization contribution in Eq.~\rf{GdeR} where it provides the dominant contribution, is not there in the double hadronic vacuum polarization contribution. The leading term in a $m_{\mu}^2$--expansion for $a_{\mu}^{({\rm HVP-C})}$ must be  $\cO(m_{\mu}^4)$ at least. In fact, a detailed analysis shows that it is $\cO\left[\frac{m_{\mu}^4}{M_{\rm H}^4}\log\left(\frac{M_{\rm H}}{m_{\mu}} \right)\right]$, with $M_{\rm H}$ a hadronic scale which in the C$\chi$QM is $M_Q$ of course.  This is the reason why the double hadronic vacuum polarization contribution to the muon anomaly is so small and, as we have shown, this is a model independent statement.

\item{\sc Comment on Radiative Corrections}

\noi
Hadronic vacuum polarization generates  part of the radiative corrections to the total $e^+ e^-$   annihilation {\it bare} cross-section into hadrons. In fact this correction leads to the following modification of the {\it bare} cross-section

{\setl
\bea
\sigma(t)_{e^+ e^-\ra {\rm hadrons}} & = & \frac{4\pi^2\alpha}{t}\ \frac{1}{\pi}\Imm\Pi^{{\rm (HVP)}}\left(t\right)\nn \\
& \Ra &\frac{4\pi^2\alpha}{t}\ 
\left[1-\ 2\ \Ree\Pi^{{\rm (HVP)}}(t)\right]\frac{1}{\pi}\Imm\Pi^{{\rm (HVP)}} \left(t\right) \,,
\eea}

\noi
which corresponds to the modification
\be
\frac{1}{\pi}\Imm\Pi^{{\rm (HVP)}} \left(t\right)\Ra
\left[1-\ 2\ \Ree\Pi^{{\rm (HVP)}}(t)\right]\frac{1}{\pi}\Imm\Pi^{{\rm (HVP)}} \left(t\right)\,,
\ee
and leads, precisely, to the muon anomaly contribution given in Eq.~\rf{rep2}.
In other words, if in the lowest order expression for the muon anomaly one inserts the {\it bare} total $e^{+} e^{-}$   annihilation cross--section into hadrons, we are indeed calculating the lowest order contribution $a_{\mu}^{({\rm HVP})}$ in Eq.~\rf{GdeR}.
This implies that the appropriate radiative corrections to the physical cross-section have been made {\it including the correction due to hadronic vacuum polarization}. The alternative is to leave the physical cross--section uncorrected for hadronic vacuum polarization, in which case, when inserted in the lowest order expression, one is then calculating: $a_{\mu}^{({\rm HVP})}+a_{\mu}^{({\rm HVP-C})}$. Then, obviously, one should not add an extra independent evaluation of $a_{\mu}^{({\rm HVP-C})}$. 

The warning here, specially for theorists, is that in using experimental hadronic cross--sections to compute hadronic vacuum polarization contributions to the muon anomaly, one should be very careful to know exactly what these cross--sections correspond to. Often the data which is used corresponds to different experiments which complicates even further the issue.

Another warning, this one for experimental physicists, concerns the dynamical constraint given in Eq.~\rf{dyncon}. In doing hadronic vacuum polarization corrections to the total cross--section numerically (e.g. involving an iterative procedure, as mentioned in some of the experimental papers) one should be careful to check that this constraint, which involves rather subtle cancellations, is indeed satisfied.      

\end{itemize}

\subsection {\normalsize Class D:
{\sc Contributions with HVP corrections at $\cO(\alpha)$}}

\begin{figure}[h]
\begin{center}
\includegraphics[width=0.4\textwidth]{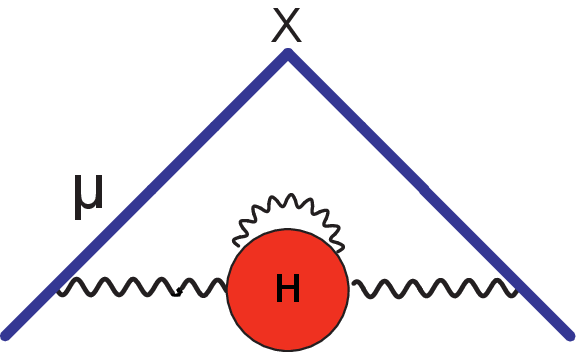}
\bf\caption{\lbl{fig:Hg}}
\vspace*{0.25cm}
{\it Feynman diagrams corresponding to the Class D contribution to the muon anomaly.
}
\end{center}
\end{figure}
\noi
In the C$\chi$QM there are two types of contributions to this class: the $\pi^0 \gamma$ exchange and the constituent quark loop with  a virtual photon insertion. They correspond to the photon propagator content illustrated in Fig.~\rf{fig:HVPgamma}.
Corresponding to these two subclasses we shall write
\be
a_{\mu}^{(\rm HVP-D)}=a_{\mu}^{(\pi^0 \gamma)}+a_{\mu}^{(Q\,, \alpha)}\,,
\ee
and discuss separately the two contributions. They are both  leading in the Large--$\Nc$ limit.
\begin{figure}[h]

\begin{center}
\includegraphics[width=14cm]{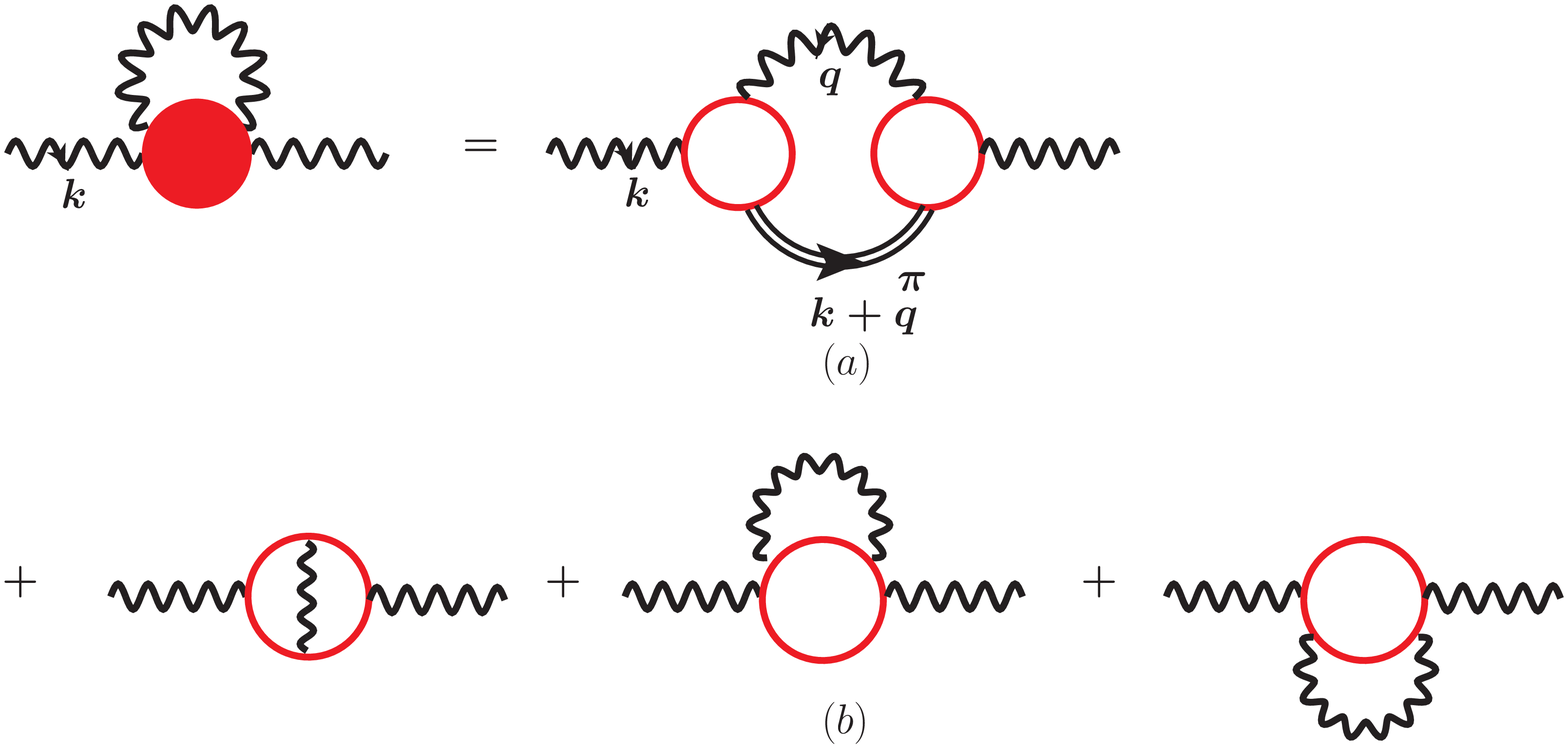}
\bf\caption{\lbl{fig:HVPgamma}}
\vspace*{0.25cm}
{\it  Feynman Diagrams which contribute to the Photon Self--Energy to $\cO(\alpha^2)$ in the C$\chi$QM.
}
\end{center}
\end{figure}

\subsubsection{\normalsize
{\sc Contribution from the $\pi^0 \gamma$ intermediate state}}

\noi
Here it is convenient  to use the representation (see page 231 in ref.~\cite{LPdeR72} and ref.~\cite{EdeR94}~\footnote{This is
a representation which is now often used by our lattice QCD colleagues to compute the hadronic vacuum polarization contribution.}):
\be\lbl{eucrep}
a_{\mu}^{(\pi^0 \gamma)}=\frac{\alpha}{\pi}\int_0^1 dz(1-z)\left[-\Pi^{(\pi^0 \gamma)}\left(-\frac{z^2}{1-z}m_{\mu}^2  \right)\right]\,,
\ee
where $\Pi^{(\pi^0 \gamma)}(k^2)$ denotes the renormalized photon self--energy from the $\pi^0 \gamma$ contribution and the integration is over the value of the self--energy in the euclidean.
In fact, we find that a better representation, which avoids renormalization issues, is the one in terms of the Adler function~\cite{PdeR}~\footnote{Fred Jegerlehner has often advocated  also the advantages of this representation.}. Using integration by parts in Eq.~\rf{eucrep} with $(1-z)=-\frac{1}{2}\frac{d}{dz}(1-z)^2$ and the fact that $\Pi(0)=0$, one finds
\be\lbl{pigamma}
a_{\mu}^{(\pi^0 \gamma)}=\frac{\alpha}{2\pi}\int_0^1 \frac{dz}{z}(1-z)(2-z)\cA^{(\pi^0 \gamma)} \left(\frac{z^2}{1-z}m_{\mu}^2 \right)\,,
\ee
with $\cA^{(\pi^0 \gamma)}(Q^2)$ the Adler function ($Q^2 =-k^2$)
\be
\cA^{(\pi^0 \gamma)}(Q^2)=-Q^2\  \frac{d\Pi^{(\pi^0 \gamma)}(k^2)}{dQ^2}\,,\quad Q^2\equiv \frac{z^2}{1-z}m_{\mu}^2\,.
\ee

In the C$\chi$QM, the $\pi^{0} \gamma\gamma$ three--point function at each vertex in the first diagram of Fig.~10(a) can be expressed in terms of the  following parametric representation:

{\setl
\bea\lbl{cqmff}
\lefteqn{\cF_{{\pi^0}^* \gamma^* \gamma^*}^{(\chi{\rm QM})}\left((k+q)^2,k^2 ,q^2 \right)=-ie^2\frac{N_c}{12\pi^2 f_{\pi}}\times }\nn \\ & & \hspace*{-1.5cm} \int_0^1 dx x  \int_0^1 dy 
\frac{2 M_Q^2}{M_Q^2-x(1-x)(1-y) k^2 -x^2 y(1-y)(k+q)^2  -xy(1-x) q^2-i\epsilon}\,.
\eea}

\noi
Here, the constituent quark mass $M_Q$ acts as an UV--regulator of the $\pi^0 \gamma$ contribution to the muon anomaly. In the limit $M_Q \ra\infty$ this form factor reduces to the $\pi^0 \gamma\gamma$ Adler, Bell--Jackiw  point--like coupling (ABJ):
\be
\cF_{{\pi^0}^* \gamma^* \gamma^*}^{(\chi{\rm QM})}\left((k+q)^2,k^2 ,q^2 \right)\xrightarrow[M_Q \ra\infty]{} -ie^2\frac{N_c}{12\pi^2 f_{\pi}}\,,
\ee
and in this limit, the contribution to the muon anomaly becomes UV--divergent. Using dimensional regularization and  the $\msb$--renormalization scheme, the result in this limit, with $m_{\mu}^2 \ll m_{\pi}^2$ for further simplification, and $\mu$ the renormalization scale, is
\be
a_{\mu}^{(\pi^0 \gamma)}(ABJ)=\left(\frac{\alpha}{\pi}\right)^3 \frac{m_{\mu}^2}{16\pi^2 f_{\pi}^2}\frac{N_c^2}{162}\left\{\log\frac{\mu^2}{m_{\pi}^2}+\frac{5}{6}+\cO\left[\frac{m_{\mu}^2}{m_{\pi}^2}\log\frac{m_{\pi}^2}{m_{\mu}^2}\right]\right\}\,.
\ee
In particular, for $\mu=M_{\rho}$, one finds
\be\lbl{ABJ}
a_{\mu}^{(\pi^0 \gamma)}(ABJ)_{\mu=M_{\rho}}=2.5\times 10^{-11}\,;
\ee
a  result which, within a $30\%$ error, is consistent with the one in ref.~\cite{BCM02}:
\be
a_{\mu}^{(\pi\gamma)}\simeq 3.7\times 10^{-11}\,,
\ee
obtained with Vector Meson Dominance like form factors:
\be\lbl{BCM}
\cF_{{\pi^0}^* \gamma^* \gamma^*}^{({\rm VMD})}\left((k+q)^2,k^2 ,q^2 \right)=
-ie^2\frac{N_c}{12\pi^2 f_{\pi}}\ 
\frac{M_{\rho}^2}{M_{\rho}^2 -k^2}\frac{M_{\rho}^2}{M_{\rho}^2 -q^2}\,.
\ee

The Feynman parameterization of the full $\pi^0 \gamma$ Adler function in  the C$\chi$QM, using the form factor expression in Eq.~\rf{cqmff},  
results in the following representation:

{\setl
\bea
\lefteqn{\cA^{(\pi^0 \gamma)}(Q^2) =  \left(\frac{\alpha}{\pi} \right)^2 N_c^2 \frac{Q^2}{16\pi^2 f_{\pi}^2}\ \frac{4}{9}\  \int_0^1 du u^2 \int_0^1 dv v \int_0^1 dw} \nn \\ & & 
 \times\int_0^1 dx x\int_0^1 dy\  \int_0^1 dx' x'\int_0^1 dy'\frac{1}{xy(1-xy)x'y'(1-x'y')}\nn \\ & & \left\{\frac{(1-y)[1-x(1-y)]}{y(1-xy)}u(1-v)+ \frac{(1-y')[1-x'(1-y')]}{y'(1-x'y')}(1-u) \right. \nn \\ & & 
 \left. + uv(1-w) -\left[uv(1-w)+u(1-v)\delta +(1-u)\delta'\right]^2\right\}\left(\frac{M_Q^2}{\cR^2}\right)^2\,,
\eea}

\noi
where 

{\setl
\bea\lbl{adlerpigamma}
\lefteqn{\hspace*{-1cm}\cR^2 = Q^2 \left\{\frac{(1-y)[1-x(1-y)]}{y(1-xy)}u(1-v)+ \frac{(1-y')[1-x'(1-y')]}{y'(1-x'y')}(1-u) \right.} \nn \\ & & 
 \left. + uv(1-w) -\left[uv(1-w)+u(1-v)\delta +(1-u)\delta'\right]^2\right\}\nn \\ & & 
 + M_Q^2 \left[\frac{u(1-v)}{xy(1-xy)}+\frac{1-u}{x'y'(1-x'y')} \right]+m_{\pi}^2 uv(1-w)\,.
\eea}

\noi
and  
\be
\delta=\frac{x(1-y)}{1-xy}\quad\annd\quad \delta'=\frac{x'(1-y')}{1-x'y'}\,.
\ee
Performing the integration over the seven Feynman parameters numerically we find
the following results:
\be\lbl{piggr}
a_{\mu}^{(\pi\gamma)}(M_Q =240~\MeV)=2.17\times 10^{-11}\,,
\ee
with a range 
\be
2.10\times 10^{-11} \le a_{\mu}^{(\pi\gamma)}\le 2.18\times 10^{-11}\,,\quad \foor\quad 230~\MeV\le M_Q \le 250~\MeV\,,
\ee
consistent with the estimate in Eq.~\rf{ABJ}.

We observe, however,  that these  results  turn out to be  an order of magnitude  smaller than the phenomenological contribution quoted in  the TABLE II of ref.~\cite{DHMZ10} (see also ref.~\cite{Hag11}) :
\be
a_{\mu}^{(\pi\gamma)}=(44.2\pm 1.9)\times 10^{-11}\,,
\ee
which uses as input the measured $\sigma(e^+ e^- \ra \pi^0 \gamma)$ cross section in the energy interval $0.60< \sqrt{s}<1.03~\GeV$~\cite{Acha03}.

\begin{itemize}
	\item {\sc Why this discrepancy?}

In order to  understand better  the underlying physics let us use instead the representation for $a_{\mu}^{(\pi\gamma)}$ in terms of the spectral function $\frac{1}{\pi}\Imm\Pi^{(\pi\gamma)}(t)$ i.e.,
\be\lbl{disprep}
a_{\mu}^{(\pi\gamma)}=\int_{m_{\pi}^2}^{\infty}
\frac{dt}{t}K\left(\frac{t}{m_{\mu}^2}\right)
\frac{1}{\pi}\Imm\Pi^{(\pi\gamma)}(t)\,.
\ee
The phenomenological determinations of $a_{\mu}^{(\pi\gamma)}$ in refs.~\cite{DHMZ10,Hag11} implicitly assume that $\frac{1}{\pi}\Imm\Pi^{(\pi\gamma)}(t)$ is completely saturated by the $\pi^0 \gamma$ intermediate state. Notice however that in the C$\chi$QM there are other intermediate states which also contribute to $\frac{1}{\pi}\Imm\Pi^{(\pi\gamma)}(t)$; they correspond to the $Q\bar{Q}$, $Q\bar{Q}\gamma$ and $Q\bar{Q}\pi$ discontinuities in the diagram (a) of Fig.~\rf{fig:HVPgamma}. These discontinuities are automatically included in the calculation of $a_{\mu}^{(\pi^0 \gamma)}$ which uses the euclidean representation in Eq.~\rf{pigamma}.  In order to compare the C$\chi$QM determination to the phenomenological ones, let us then  restrict $\frac{1}{\pi}\Imm\Pi^{(\pi\gamma)}(t)$ to the contribution from the on--shell $\pi^0 \gamma$ intermediate state only. Then
\be\lbl{spcpig}
\frac{1}{\pi}\Imm\Pi^{(\pi\gamma)}(t)\big\vert_{\pi\gamma}=\left(\frac{\alpha}{\pi} \right)^2 \frac{N_c^2}{16\pi^2 f_{\pi}^2}\frac{1}{54}\ t\ \left(1-\frac{m_{\pi}^2}{t} \right)^3 \ \Big\vert {\tilde{\cF}}_{{\pi^0} \gamma^* \gamma}^{(\chi{\rm QM})}\left(m_{\pi}^2,t ,0 \right)\Big\vert^2\,,
\ee
with
\be\lbl{cqmffos}
\hspace*{-0.25cm}
{\tilde{\cF}}_{{\pi^0} \gamma^* \gamma}^{(\chi{\rm QM})}\left(m_{\pi}^2,t ,0 \right)
= \int_0^1 dx x  \int_0^1 dy 
\frac{2 M_Q^2}{M_Q^2-x(1-x)(1-y) t -x^2 y(1-y)m_{\pi}^2 -i\epsilon}\,.
\ee
This form factor has an imaginary part which, for $m_{\pi}\leq 2M_{Q}$, is:
\be\lbl{imag}
\frac{1}{\pi}\Imm{{\tilde\cF}}^{(\chi QM)}_{\pi^{0}\gamma^* \gamma} (m^2_\pi, t,0) = \frac{4M_Q^2}{t-m_\pi^2}\log\frac{\sqrt{t}}{2M_Q}\left( 1 + \sqrt{1-\frac{4 M_{Q}^2}{t}}\right) \theta(t-4 M_{Q}^2)\,.
\ee
and a real part:

{\setl
\bea
\hspace*{-1cm}
\Ree{\tilde{\cF}^{(\chi QM)}_{\pi^0\gamma^* \gamma} (m^2_\pi, t,0)} & = & \frac{4M_Q^2}{t-m^2_\pi}\left[ 
\arctan^2\left( \frac{1}{\sqrt{\frac{4M_Q^2}{m^2_\pi}-1}} \right) 
\right.\nonumber \\
&  + & \left. \begin{cases}
-\arctan^2\left( \frac{1}{\sqrt{\frac{4M_Q^2}{t}-1}} \right)\,,\qquad m_{\pi}^2 \leq t \leq 4M_Q^2\\
\ln^2  \frac{\sqrt{t}}{2M_Q}\left(1 + \sqrt{1-\frac{4M_Q^2}{t}}\right) -  \frac{\pi^2}{4}\,,\quad  t\geq 4M_Q^2 
\end{cases}\right]\,.
\eea}

\noi
The shape of the spectral function in Eq.~\rf{spcpig}, in units of $\left(\frac{\alpha}{\pi}\right)^2$ and for the value $M_Q =240~\MeV$, is shown in Fig.~\rf{fig:Specpig}.

\begin{figure}[h]
\begin{center}
\includegraphics[width=0.75\textwidth]{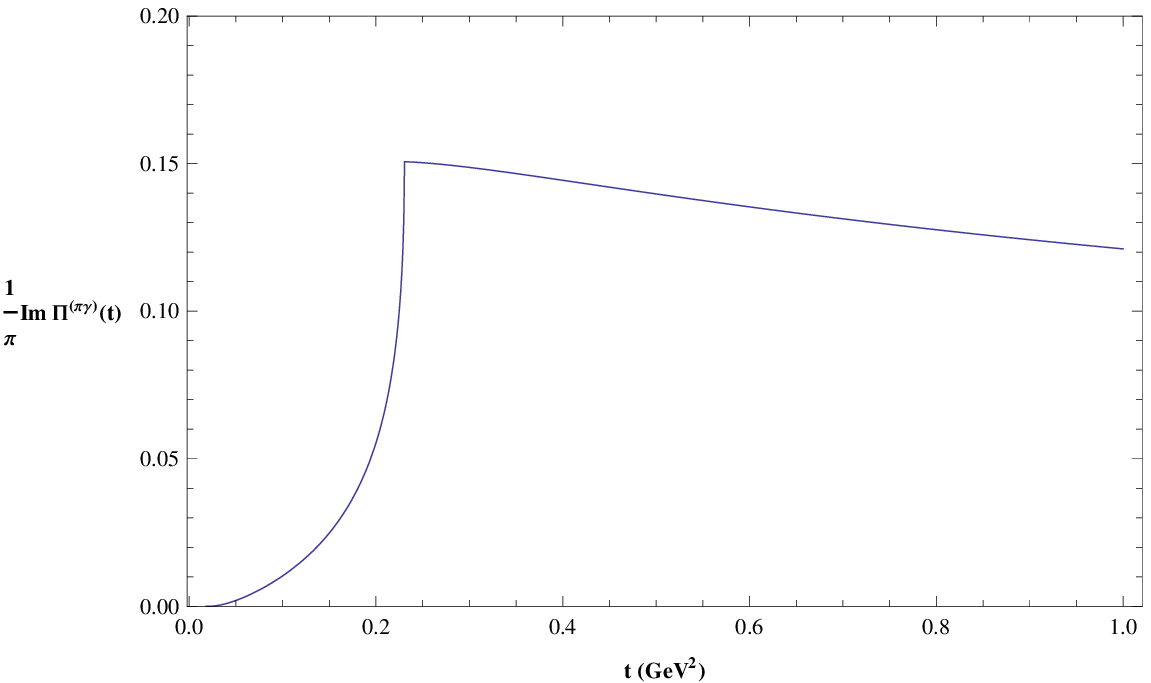}
\bf\caption{\lbl{fig:Specpig}}
\vspace*{0.25cm}
{\it  The Spectral Function in Eq.~\rf{spcpig}, in units of $\left(\frac{\alpha}{\pi}\right)^2$, for $M_Q =240~\MeV$. 
}
\end{center}
\end{figure}

The result of the integral in Eq.~\rf{spcpig} with the spectral function plotted in Fig.~9 is then
\be
a_{\mu}^{(\pi\gamma)}({\rm Eq.}~\rf{disprep}) = 3.0\times 10^{-11} \quad \foor\quad M_Q = 240~\MeV\,,
\ee
a result which is slightly higher than the full C$\chi$QM contribution in Eq.~\rf{piggr} but still well below the phenomenological determinations~\cite{Hag11,DHMZ10}. 

Let us then try to simplify the phenomenological determination as much as possible to see where the big contribution comes from. For that, it will be sufficient to approximate  Eq.~\rf{disprep}
as follows:	
\be
a_{\mu}^{(\pi\gamma)}\sim\left(\frac{\alpha}{\pi} \right)\int_{m_{\pi}^2}^{\infty}
\frac{dt}{t}\frac{1}{3}\frac{m_{\mu}^2}{t}
\frac{1}{\pi}\Imm\Pi^{(\pi\gamma)}(t)\,,
\ee
and use a narrow width expression for the spectral function, which as we shall soon see, is dominated by the $\omega$ contribution. This results in the simple formula:
\be\lbl{simple}
a_{\mu}^{(\pi\gamma)}\sim\frac{1}{3}\frac{m_{\mu}^2}{M_{\omega}^2}\frac{4}{\pi}\frac{\Gamma(\omega\ra e^+e^-)}{M_{\omega}}\frac{\Gamma(\omega\ra \pi^0 \gamma)}{\Gamma_{\omega}}=53\times 10^{-11}\,,
\ee
which reproduces, in order of magnitude, the phenomenological estimates. We can, therefore, see that the {\it big number}  comes from the large  experimental value of the branching ratio 
\be\lbl{omega}
\frac{\Gamma(\omega\ra \pi^0 \gamma)}{\Gamma_{\omega}}\simeq 8\times 10^{-2}\,.
\ee
Notice that  in the case of the $\rho$ contribution the corresponding branching ratio is much smaller:
\be
\frac{\Gamma(\rho\ra \pi^0 \gamma)}{\Gamma_{\rho}}\simeq 6\times 10^{-4}\,.
\ee
It is the large branching ratio in Eq.~\rf{omega}
which the C$\chi$QM fails to reproduce!

Phenomenologically, the large branching ratio $\frac{\Gamma(\omega\ra \pi^0 \gamma)}{\Gamma(\rho\ra \pi^0 \gamma)}$  is due to the $\omega$--$\phi$ mixing and the fact that the $\phi$ is an almost pure $s\bar{s}$ state~\footnote{We thank Marc Knecht for reminding us of this.}. By construction, the C$\chi$QM form factor is $SU(3)$ invariant and, therefore, like any model which is $SU(3)$ invariant, fails to reproduce this phenomenological fact. 
\end{itemize}

We are aware of the fact that in the C$\chi$QM there are also further contributions of the $\pi^0 \gamma$ subclass: those from the $\eta_8 \gamma$ and $\eta_0 \gamma$ intermediate states. We refrain from discussing them because their comparison with their corresponding phenomenological determinations requires issues like $\eta_8 - \eta_0$ mixing as well as the question of the $\eta'$ mass in Large--$\Nc$ which are beyond the scope of the model we are discussing.

\subsubsection
{\sc Contribution from the Quark Loop to $\cO(\alpha)$}

\noi
This contribution is given by the following integral representation:
\be\lbl{D}
a_{\mu}^{(Q\,,\alpha)}=\int_{4M_{Q}^2}^{\infty}
\frac{dt}{t}K(t/m_{\mu}^2)\ 
\frac{1}{\pi}\Imm\Pi^{(Q\,,\alpha)}(t)\,,
\ee
where
\be\lbl{CQMalpha}
\frac{1}{\pi}\Imm\Pi^{(Q\,,\alpha)}(t)  =  	\left(\frac{\alpha}{\pi}\right)^2 \ N_c \left[\left(\frac{2}{3}\right)^4+\left(-\frac{1}{3}\right)^4 +\left(-\frac{1}{3}\right)^4\right]\ \rho_{\rm KS}(t)\,,
\ee
with $\rho_{\rm KS}(t)$ given in Eq.~\rf{KS}. As $M_Q \ra\infty$ it decouples with a leading behaviour:

{\setl
\bea
a_{\mu}^{(Q\,,\alpha)} & \simeq & \left(\frac{\alpha}{\pi}\right)^3 N_c \ \frac{2}{9}\ 
\frac{1}{3}\frac{m_{\mu}^2}{M_Q^2}\left(\int_{4 M_{Q}^2}^{\infty} \frac{dt}{t}\frac{M_{Q}^2}{t}\ \rho_{\rm KS}(t)=\frac{41}{162}\right)\nn \\
 & = &  13.7\times 10^{-11} \quad \foor\quad M_Q =240~\MeV\,.
\eea}

\noi
The full numerical evaluation gives 
\be
a_{\mu}^{(Q\,,\alpha)}(M_Q =240~\MeV)=11.4\times 10^{-11}\,,
\ee
with a range 
\be
10.6\times 10^{-11} \le a_{\mu}^{(Q\,,\alpha)}\le 12.3\times 10^{-11}\,,\quad \foor\quad 250~\MeV\ge M_Q \ge 230~\MeV\,.
\ee

Altogether we find that, although the $\pi^0 \gamma$ exchange contribution increases logarithmically as a function of $M_Q$, while the quark loop decouples as an inverse power of $M_Q^2$, their ratio for $M_Q^2$ large goes as 
\be
\frac{a_{\mu}^{(\pi\gamma)}}{a_{\mu}^{(Q\,,\alpha)}}\Big\vert_{M_Q\ra\infty}\sim\  \Nc\frac{ M_{Q}^2}{16\pi^2 f_{\pi}^2}\  \frac{27}{82} \log\frac{M_{Q}^2}{m_{\pi}^2} 
\ee
and, therefore,  
for values of the constituent quark mass in the range $250~\MeV\ge M_Q \ge 230~\MeV$, it is the quark loop contribution which still dominates.

The total  sum of the two contributions of Class {\bf D} in the C$\chi$QM is then:
\be
a_{\mu}^{(\rm HVP-D)}(M_Q =240~\MeV)=13.6\times 10^{-11}\,,
\ee
with a range for this total
\be
12.8\times 10^{-11} \le a_{\mu}^{(\rm HVP-D)}\le 14.4\times 10^{-11}\,,\quad \foor\quad 250~\MeV\ge M_Q \ge 230~\MeV\,.
\ee
Except for the $\pi^0 \gamma$  contribution, it is difficult to compare the overall C$\chi$QM prediction for the {\bf Class D} contributions with the phenomenological estimates. The reason is  that, a priori, when inserting a physical observable to evaluate the diagram in Fig.~\rf{fig:Hg} one needs two types of contributions: the one from the cross section $\sigma(e^+ e^- \ra {\rm Hadrons} + \gamma)$ and the one from the interference of the amplitude $e^+ e^- \ra {\rm Hadrons}$ with  the same amplitude where a virtual photon has been emitted and reabsorbed. In fact, individually, these two contributions are infrared divergent, which complicates things even more. This is a place where it would be interesting to see if lattice QCD can eventually make an estimate of these {\bf Class D} contributions which, so far, remain poorly known phenomenologically. 

Table~1 gives a summary of the results for the four classes of contributions discussed here and evaluated within the framework of the C$\chi$QM, with a total
\be
a_{\mu}^{[\rm HVP-(A,B,C,D)]}= \left(-64 \pm 12\right) \times 10^{-11}\,.  
\ee

This C$\chi$QM result is to be compared with the number quoted in the latest evaluation in ref.~\cite{Hagietal11} for this contribution which,  however, does not include the important contributions from the $\pi^0 \gamma$ and $\eta \gamma$ intermediate states already incorporated in the lowest order HVP contribution:
\be
a_{\mu}^{[\rm HVP-next order]}(e^+  e^-) = (-98.4\pm 0.6_{\rm exp} \pm 0.4_{\rm rad})
\times 10^{-11}\,.
\ee
Again, except for the $\pi^0 \gamma$ issue already discussed, the agreement within the errors of the model is quite reasonable.

\begin{table*}[h]
\caption[Results]{{\it Results  for the HVP contributions of $\cO\left(\frac{\alpha}{\pi} \right)^3$ in the C$\chi$QM.}}

\lbl{table1}
\begin{center}
\begin{tabular}{|c|c|} \hline \hline  {\bf Class} &
{\bf Result in $10^{-11}$ units}\\
 \hline \hline
 A  & $ -171\pm 10$
\\  B & $89\pm 7$\\
 C & $ 2.2\pm 0.3$ \\
  D($\pi \gamma$) & $2.2 \pm 0.1 $ \\ 
 D(Q--loop) & $13.5\pm 0.5$\\
\hline 
Total & $ -64 \pm 12$ \\
\hline\hline
\end{tabular}
\end{center}
\end{table*} 

\noi
\vspace*{1cm}
\section{\normalsize Hadronic Light--by--Light Scattering  Contributions.}
\setcounter{equation}{0}
\def\theequation{\arabic{section}.\arabic{equation}}
\noi
The standard representation of the contribution to the muon anomaly
from the hadronic light--by--light scattering shown in Fig.~\rf{fig:LbyL1} is given by the integral~\cite{ABDK69}:

\begin{figure}[h]

\begin{center}
\includegraphics[width=0.6\textwidth]{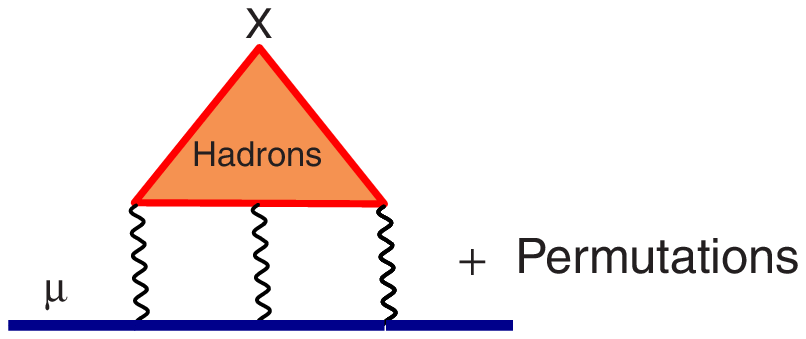}
\bf\caption{\lbl{fig:LbyL1}}
\vspace*{0.25cm}
{\it  Hadronic Light--by--Light Scattering contribution to the Muon Anomaly.
}
\end{center}
\end{figure}

{\setl
\bea
\lbl{aH}
a^{(\rm HLbyL)}_{\mu} &=& \frac{-ie^{6}}{48m_{\mu}}\int\!\frac{d^4 q_1}{(2\pi)^4}\!\!\int\frac{d^4 q_2}{(2\pi)^4}\frac{1}{q_1^2 q_2^2 (q_1 +q_2)^2}
\left[\frac{\partial}{\partial q^{\mu}}\Pi_{\lambda\nu\rho\sigma}^{({\rm H})}(q,q_1, q_3, q_2)\right]_{q=0}
 \nn \\ 
 & &\hspace*{-1cm}\times \,\tr\left\{
        (\pslsh + m_{\mu})[\gamma^{\mu},\gamma^{\lambda}](\pslsh +m_{\mu}) \gamma^{\nu}\frac{1}{\pslsh+\qsls_2-m_{\mu}}\gamma^{\rho}\frac{1}{\pslsh -\qsls_1-m_{\mu})}\gamma^{\sigma}\right\}\,,
\eea}

\noi
where $\Pi_{\mu\nu\rho\sigma}^{({\rm H})}(q,q_1, q_3, q_2)$, with $q=p_{2}-p_{1}=-q_1 -q_2 -q_3$, denotes the off--shell photon--photon scattering amplitude induced by hadrons,

{\setl
\bea
\Pi_{\mu\nu\rho\sigma}^{({\rm H})}(q,q_1, q_3, q_2)	& = & 
\!\!\int\!\! d^4x_1\!\!\int\!\! d^4x_2\!\!\int\!\! d^4x_3\  \exp[{-i(q_1\cdot x_1\! +\!q_2\cdot x_2\!+\!q_3\cdot x_3)]} 
\nn \\
 &  & \times\langle 0 \vert T\{J_{\mu}(0)\,, J_{\nu}(x_1)\,, J_{\rho}(x_2)\,,  J_{\sigma}(x_3)\}\vert 0\rangle\,,
\eea}

\noi
and  $J_{\mu}(x)=\sum_{q}Q_{q}\bar q(x) \gamma_{\mu}q(x)$ is  the Standard Model electromagnetic hadronic current where, for the light quarks, $Q_{q}={\rm diag} (2/3,-1/3,-1/3)$.

In the C$\chi$QM there are two types of contributions: the Constituent Quark Loop (CQL) contribution shown in Fig.~\rf{fig:LxL-Loop} 
\begin{figure}[h]
\begin{center}
\includegraphics[width=0.5\textwidth]{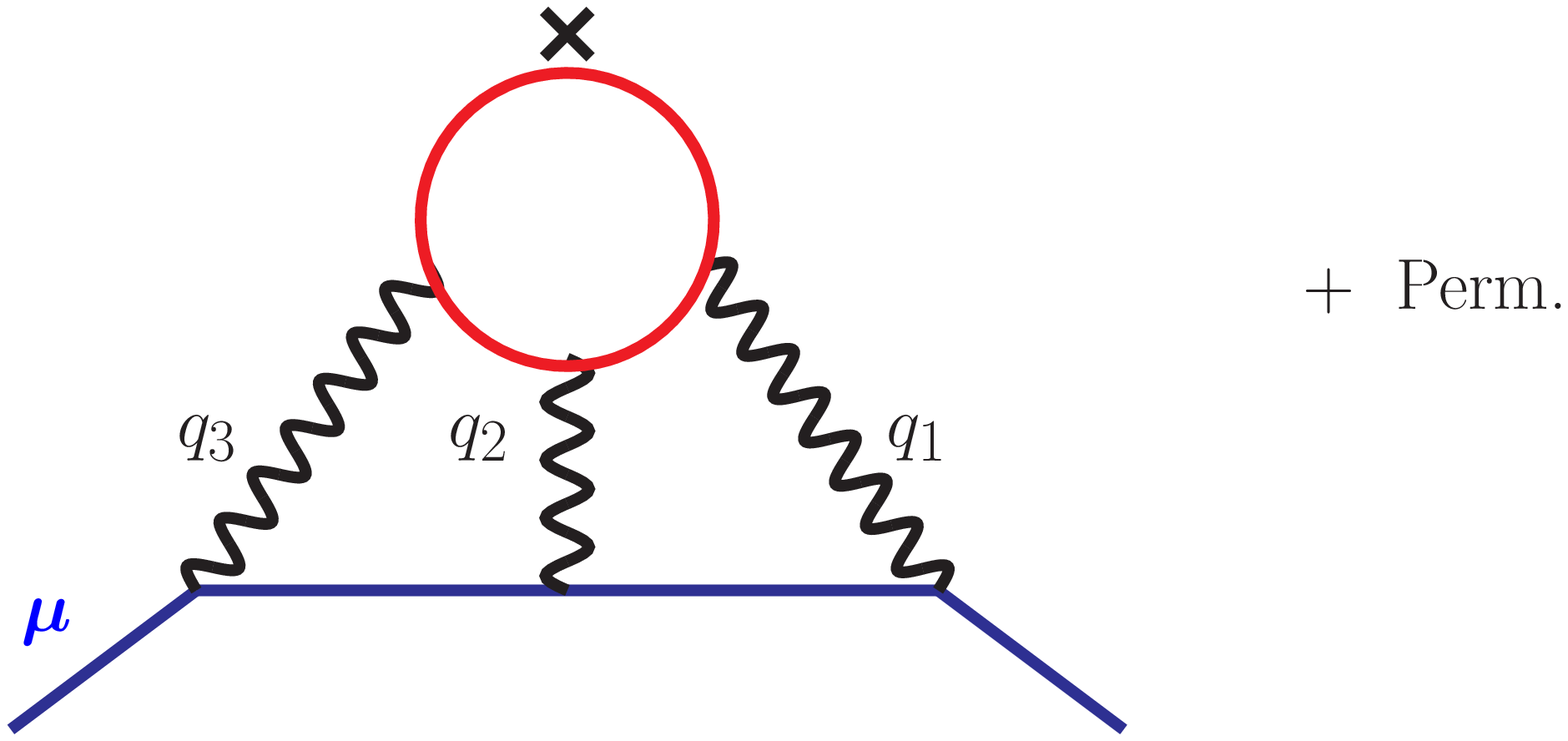}
\bf\caption{\lbl{fig:LxL-Loop}}
\vspace*{0.25cm}
{\it  Constituent Quark Loop Contribution to the Muon Anomaly in the C$\chi$QM.
}
\end{center}
\end{figure}
and the Goldstone Exchange Contribution shown in Fig.~\rf{fig:LxL-pi} with constituent quark loops at each vertex.
\begin{figure}[h]
\begin{center}
\includegraphics[width=0.8\textwidth]{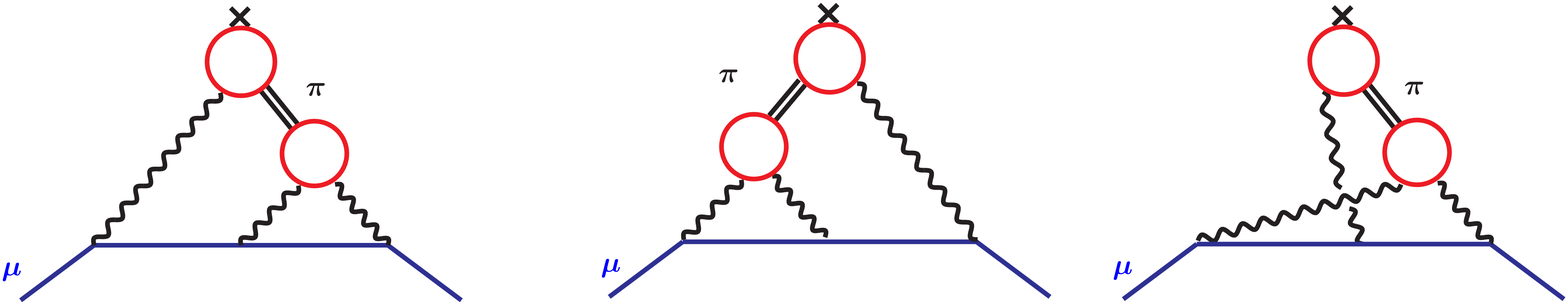}
\bf\caption{\lbl{fig:LxL-pi}}
\vspace*{0.25cm}
{\it  Goldstone Exchange Contribution to the Muon Anomaly in the C$\chi$QM.}
\end{center}
\end{figure}
We shall consider these two types of contributions, both leading in the $1/\Nc$--expansion, separately. .

\subsection {\normalsize Class A: 
	{\sc The Constituent Quark Loop Contribution.}}

\noi
This contribution can be obtained from the QED analytic calculation of Laporta and Remmidi~\cite{LR93} with the result

{\setl
\bea\lbl{LbyLCQL}
a_{\mu}^{\rm (HLbyL)}({\rm CQL}) &  = & \left(\frac{\alpha}{\pi}\right)^3 N_c \left(\sum_{q=u,d,s} Q_q^4 \right)\left\{
\left[\frac{3}{2}\zeta(3)-\frac{19}{16} \right] \frac{m_{\mu}^2}{M_{Q}^2}\right. \nn \\
 &- &  \frac{m_{\mu}^4}{M_{Q}^4}\left[\frac{161}{3240}\log^2 \frac{M_{Q}^2}{m_{\mu}^2}+\frac{16189}{97200}\log \frac{M_{Q}^2}{m_{\mu}^2} -\frac{13}{18}\zeta(3)+\frac{161}{1620}\frac{\pi^2}{6}+\frac{831931}{972000}\right]
 \nn \\
& + &  
 \left.\cO\left(\frac{m_{\mu}^6}{M_{Q}^6}\log^2 \frac{m_{\mu}^2}{M_{Q}^2} \right)
\right\}\,.
\eea}

\noi
A plot of this contribution versus $M_Q$ is shown in Fig.~\rf{fig:LbyL-CQL}. We find 
\be
a_{\mu}^{\rm (HLbyL)}({\rm CQL})=82.2\times 10^{-11}\quad {\rm at}\quad M_Q=240~\MeV \,,
\ee

\begin{figure}[h]

\begin{center}
\includegraphics[width=0.8\textwidth]{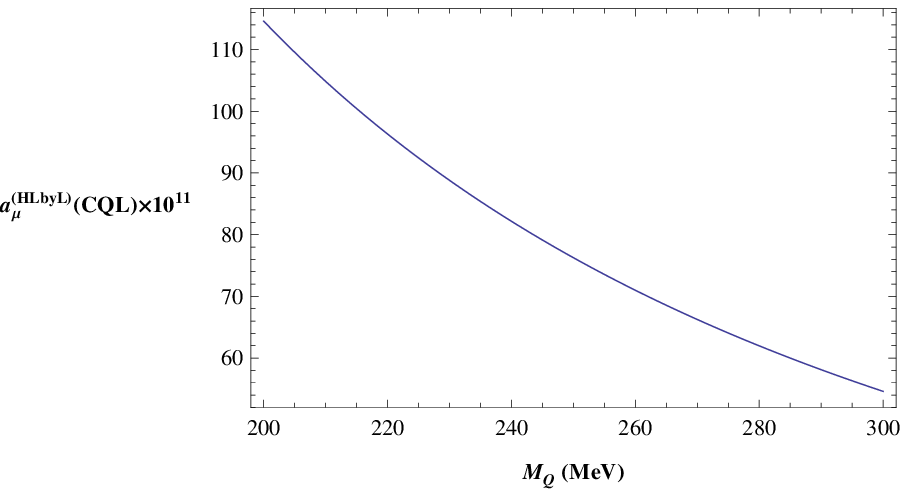}
\bf\caption{\lbl{fig:LbyL-CQL}}
\vspace*{0.25cm}
{\it   $a_{\mu}^{\rm (HLbyL)}({\rm CQL})$ in Eq.~\rf{LbyLCQL} in $10^{-11}$ units.}
\end{center}
\end{figure}

\noi
with a range
\be
76.3\times 10^{-11}\leq a_{\mu}^{\rm (HLbyL)}({\rm CQL})\leq 88.8\times 10^{-11}\quad\foor\quad 250~\MeV \geq M_Q \geq 230~\MeV\,.
\ee

The gluonic corrections of $\cO\left(\frac{\als}{\pi} \right)$ to the leading term in Eq.~\rf{LbyLCQL} have been recently calculated in ref.~\cite{BM11} and found to be rather small.

\subsection {\normalsize Class B: 
	{\sc The $\pi^0$ Exchange Contribution.}}

\noi
The expression for this contribution in terms of the vertex form factors 
$\cF_{{\pi^0}^* \gamma^* \gamma^*}$ and $\cF_{{\pi^0}^* \gamma \gamma^*}$
can be found in ref.~\cite{KN02}. When applied to the C$\chi$QM we have:

{\setl
\bea\lbl{LbyGCQL}
\hspace*{-1.5cm}
a_{\mu}^{\rm (HLbyL)}(\pi^0)_{\chi {\rm QM}} & = & e^2 \ \frac{8}{3} \int\frac{d^4 q_1}{(2\pi)^4}\int\frac{d^4 q_2}{(2\pi)^4}\frac{1}{q_1^2 q_2^2 (q_1 +q_2)^2 [q_{1}^2 +2 p\cdot q_1][q_{2}^2 -2 p\cdot q_2]}\nn \\
 & \times & \left[ T_1 (q_1,q_2;p)\frac{\cF_{{\pi^0}^* \gamma^* \gamma^*}^{(\chi{\rm QM})}\left(q_{2}^2,q_{1}^2 ,(q_{1}+q_{2})^2 \right)\cF_{{\pi^0}^* \gamma^* \gamma^*}^{(\chi{\rm QM})}\left(q_{2}^2,q_{2}^2 ,0 \right)}{q_{2}^2 -m_{\pi}^2}\right. \nn \\
  & + & \left. { T_2 (q_1,q_2;p)\frac{\cF_{{\pi^0}^* \gamma^* \gamma^*}^{(\chi{\rm QM})}\left((q_{1}+q_{2})^2,q_{1}^2 ,q_{2}^2 \right)\cF_{{\pi^0}^* \gamma^* \gamma^*}^{(\chi{\rm QM})}\left((q_{1}+q_{2})^2,(q_{1}+q_{2})^2 ,0 \right)}{(q_{1}+q_{2})^2 -m_{\pi}^2}}\right]\,,  
\eea}
 
\noi
where

{\setl
\bea
T_1 (q_1 ,q_2; p) & = & 2(p\cdot q_1)(p\cdot q_2)(q_1\cdot q_2)-2
(p\cdot q_2)^2 q_1 ^2 -(p\cdot q_1 )(q_1 \cdot q_2)q_{2}^2\nn \\
 & + & 3(p\cdot q_2) q_{1}^2 q_{2}^2 -2(p\cdot q_2 )(q_1 \cdot q_2)^2 + 2m_{\mu}^2 \left[q_1 ^2 q_2 ^2 -(q_1 \cdot q_2 )^2\right]
\eea}

\noi 
originates in the first and second diagrams of Fig.~\rf{fig:LxL-pi}, which give identical contributions, while

{\setl
\bea
T_2 (q_1 ,q_2; p) & = & 2(p\cdot q_1)(p\cdot q_2)(q_1\cdot q_2)-2
(p\cdot q_1)^2 q_2 ^2 +(p\cdot q_1 )(q_1 \cdot q_2)q_{2}^2\nn \\
 & + & (p\cdot q_1) q_{1}^2 q_{2}^2  + m_{\mu}^2 \left[q_1 ^2 q_2 ^2 -(q_1 \cdot q_2 )^2\right]
\eea}

\noi
originates in the third diagram of Fig.~\rf{fig:LxL-pi}.

It is well known~\cite{KNPdeR02} that, asymptotically for $M_Q \gg m_{\pi}$, the $\pi^0$--exchange  contribution must behave as: 

\be
\lbl{piefft}
a_{\mu}^{\rm (HLbyL)}(\pi^0)_{\chi {\rm QM}} =\Big(\frac{\alpha}{\pi}\Big)^{3}N_{c}^2 \frac{m_{\mu}^{2}}{16\pi^2 f_{\pi}^{2}}
\Big[\frac{1}{3} \ln^2\!\frac{M_Q}{m_{\pi}}  +
\cO\Big(\ln\frac{M_{Q}}{m_{\pi}}\Big)+\cO({\rm cte.})\Big]\,.
\ee
In the C$\chi$QM  this contribution can be evaluated exactly. Notice that  $\cF_{{\pi^0}^* \gamma^* \gamma^*}^{(\chi{\rm QM})}\left(q^2,q^2 ,0 \right)$ has a simple analytic expression:

{\setl
\bea\lbl{QQvertex}
\cF_{{\pi^0}^* \gamma^* \gamma^*}^{(\chi{\rm QM})}\left(q^2,q^2 ,0 \right) & = & -ie^2
\frac{\Nc}{12\pi^2 f_{\pi}}\int_0^1 dx \frac{M_Q^2}{M_Q^2 -x(1-x)q^2}\nn \\
& = & ie^2
\frac{\Nc}{12\pi^2 f_{\pi}}\left[\frac{M_{Q}^2}{Q^2}
\frac{2}{\sqrt{1+\frac{4M_{Q}^2}{Q^2}}}
\log{\frac{\sqrt{1+\frac{4M_{Q}^2}{Q^2}}-1}
{\sqrt{1+\frac{4M_{Q}^2}{Q^2}}+1}}\right]\nn \\
 & = & -ie^2
\frac{\Nc}{12\pi^2 f_{\pi}}\frac{1}{2\pi i}\int_{c_s -i\infty}^{c_s +i\infty} ds \left(\frac{q^2}{[-M_{Q}^2]} \right)^{-s}\ \frac{\Gamma(s)\Gamma(1-s)^3}{\Gamma(2-2s)}\,,
\eea}

\noi
where the expression in the third line gives a very useful representation for analytic evaluations. The full expression of the  other vertex function is:

{\setl
\bea
\lefteqn{\cF_{{\pi^0}^* \gamma^* \gamma^*}^{(\chi{\rm QM})}\left(q_2^2,q_1^2 ,q_3^2 \right)=-ie^2 \frac{N_c}{12\pi^2 f_{\pi}}\times} \nn \\
& & \int_0^1 dx x  \int_0^1 dy 
\frac{2 M_Q^2}{M_Q^2-x(1-x)(1-y) q_1^2 -x^2 y(1-y)q_2^2  -xy(1-x) q_3^2}\,,
\eea}

\noi
for which the following representation can also be used:

{\setl
\bea
\lefteqn{\hspace*{3cm}\cF_{{\pi^0}^* \gamma^* \gamma^*}^{(\chi{\rm QM})}\left(q_2^2,q_1^2 ,q_3^2 \right)= -ie^2 \frac{N_c}{12\pi^2 f_{\pi}}\times}\nn \\ & &
\left(\frac{1}{2\pi i}\right)^3\ 2 
\int\limits_{c_1-i\infty}^{c_1+i\infty} ds_1\ \left(\frac{q_1^2}{[-M_Q^2]}\right)^{-s_1}
\int\limits_{c_2-i\infty}^{c_2+i\infty} ds_2\  \left(\frac{q_2^2}{[-M_Q^2]}\right)^{-s_2}
\int\limits_{c_3-i\infty}^{c_3+i\infty} ds_3\left(\frac{q_3^2}{[-M_Q^2]}\right)^{-s_3}\nn \\ & & 
\times\frac{\Gamma(1-s_1 -s_2)\Gamma(1-s_1 -s_3)\Gamma(1-s_2 -s_3)}{\Gamma(3-2 s_1 -2 s_2 -2 s_3)}
\Gamma(s_1)\Gamma(s_2)\Gamma(s_3) \Gamma(1-s_1 -s_2 -s_3)\,.
\eea}

\noi
The interest of this Mellin--Barnes representation is that it only modifies the powers of the propagators: $q_1^2$, $q_2^2$ and $(q_1 +q_2)^2$ in the first line of Eq.~\rf{LbyGCQL}, and it provides a systematic way~\footnote{See e.g.  ref.~\cite{AGdeR08} for an example of this method.} to compute the asymptotic expansion in $\frac{m_{\mu}^2}{M_Q^2}$ and $\frac{m_{\mu}^2}{m_{\pi}^2}$ powers, and powers of logarithms. We postpone, however, this analytic calculation to a forthcoming publication and, instead,  proceed  here to a numerical evaluation.

In order to evaluate $a_{\mu}^{\rm HLbyL}(\pi^0)_{\chi {\rm QM}}$ in Eq.~\rf{LbyGCQL} numerically, it is useful to apply to the integrand in that equation the technique of Gegenbauer polynomial expansion, as was done in ref.~\cite{KN02}. Then one can reduce the $q_1$ and $q_2$ integrations  to two euclidean integrals over $Q_1 ^2 \equiv -q_{1}^2$ and $Q_2 ^2 \equiv -q_{2}^2$ ( both from $0$ to $\infty$), and an integral over  $\cos\theta$  with $\theta$ the angle between the two euclidean four--vectors $Q_1$ and $Q_2$. The integrand in question, which is explicitly given in ref.~\cite{JN09}, is then very convenient for numerical integration.

We find
\be a_{\mu}^{\rm (HLbyL)}(\pi^0)_{\chi {\rm QM}} =68.0\times 10^{-11} \quad \foor \quad M_Q=240~\MeV\,,
\ee
with a range
\be a_{\mu}^{\rm (HLbyL)}(\pi^0)_{\chi {\rm QM}}  =64.6\times 10^{-11} \quad \foor \quad M_Q=230~\MeV\,,
\ee
and
\be a_{\mu}^{\rm (HLbyL)}(\pi^0)_{\chi {\rm QM}} =71.3\times 10^{-11} \quad \foor \quad M_Q=250~\MeV\,.
\ee
The result
\be\lbl{pilbyl}
a_{\mu}^{\rm (HLbyL)}(\pi^0)_{\chi {\rm QM}} = (68\pm 3)\times 10^{-11}\,,
\ee
which does not include the systematic error of the model,
agrees well with the phenomenological determinations of this contribution which, according to the most recent update~\cite{Nyetal12} and depending on the underlying phenomenological model for the form factors $\cF_{{\pi^0}^* \gamma^* \gamma^*}^{(\chi{\rm QM})}\left(q_2^2,q_1^2 ,q_3^2 \right)$ vary between 
\be
a_{\mu}^{\rm (HLbyL)}(\pi^0)_{\rm phen.}=(57.4\pm 4.6)\times 10^{-11}
\quad\annd\quad
a_{\mu}^{\rm (HLbyL)}(\pi^0)_{\rm phen.}=(80.1\pm 4.7)\times 10^{-11}\,.
\ee

Again, for the same reasons mentioned at the end of Section III.4.1, we do not discuss here the contributions from the $\eta$ and $\eta'$ exchanges.

It is a fact that asymptotically, for $M_Q \ra \infty$, the $\pi^0$ contribution largely dominates the  Constituent Quark Loop contribution:
\be
\frac{a_{\mu}^{\rm (HLbyL)}(\pi^0)}{a_{\mu}^{\rm (HLbyL)}({\rm CQL})}\Big\vert_{M_Q \ra\infty} \sim
\frac{N_c M_Q^2}{16\pi^2 f_{\pi}^2}\ \frac{1}{\zeta(3)-\frac{19}{24}}\ \log^2 \frac{M_Q}{m_{\pi}}\,;
\ee
however, this asymptotic behaviour is far from being reached at values of $M_Q$ between  $230~\MeV$ and $250~\MeV$, for which the Constituent Quark Loop contribution still dominates over the Goldstone contribution.

For the total hadronic light--by--light contribution in the C$\chi$QM, which includes the quark loop contribution as well as the $\pi^0$--exchange contributions, we then find
\be\lbl{lbylt}
148\times 10^{-11}\le a^{\rm (HLbyL)}_{\mu} ({\rm C} \chi {\rm QM})\leq 153\times 10^{-11}\,,\quad \foor\quad 250~\MeV \geq M_Q \geq 230~\MeV\,. 
\ee
This result, which does not include the systematic error of the model, has 
to be compared with the phenomenological estimate 
\be\lbl{plbyl}
a^{(\rm HLbyL)}=(122\pm 18)\times 10^{-11}\,,
\ee
for the total of the hadronic contributions not suppressed in the $1/\Nc$--expansion (see e.g. ref.~\cite{PdeRV10} for details). Within the expected systematic uncertainties they compare rather well.

The interesting feature which emerges from this calculation  is the observed balance between the Goldstone contribution and the Quark Loop contribution. Indeed, as the constituent quark mass $M_Q$ gets larger and larger, the Goldstone contribution dominates; while for $M_Q$ smaller and smaller  it is the Quark Loop contribution which dominates. This is illustrated by the plot of the total $a^{(\rm HLbyL)}_{\mu} ({\rm C} \chi {\rm QM})$ versus $M_Q$ shown in Fig.~\rf{fig:LbyLtotal}. What this plot shows is in flagrant contradiction with the results reported in ref.~\cite{GFW11} based in a calculation using a Dyson--Schwinger inspired model. In this model, the authors find a contribution from the $\pi^0$--exchange which, within errors, is compatible with the other phenomenological determinations and, in particular, with our C$\chi$QM result in Eq.~\rf{pilbyl}; yet their result for the equivalent contribution to the quark loop turns out to be almost twice as large with a total contribution
\be\lbl{germans}
a^{(\rm HLbyL)}({\rm ref.}~\text{\cite{GFW11}})=(217\pm 91)\times 10^{-11}\,.
\ee
The central value of this result would require a ridicously small value of $M_Q$ in order to be reproduced by the C$\chi$QM and, furthermore, for such a small value of $M_Q$ the $\pi^0$--exchange contribution would be far too small as compared to all the phenomenological estimates, including the one in ref.~\cite{GFW11}.  
\begin{figure}[h]
\begin{center}
\includegraphics[width=0.8\textwidth]{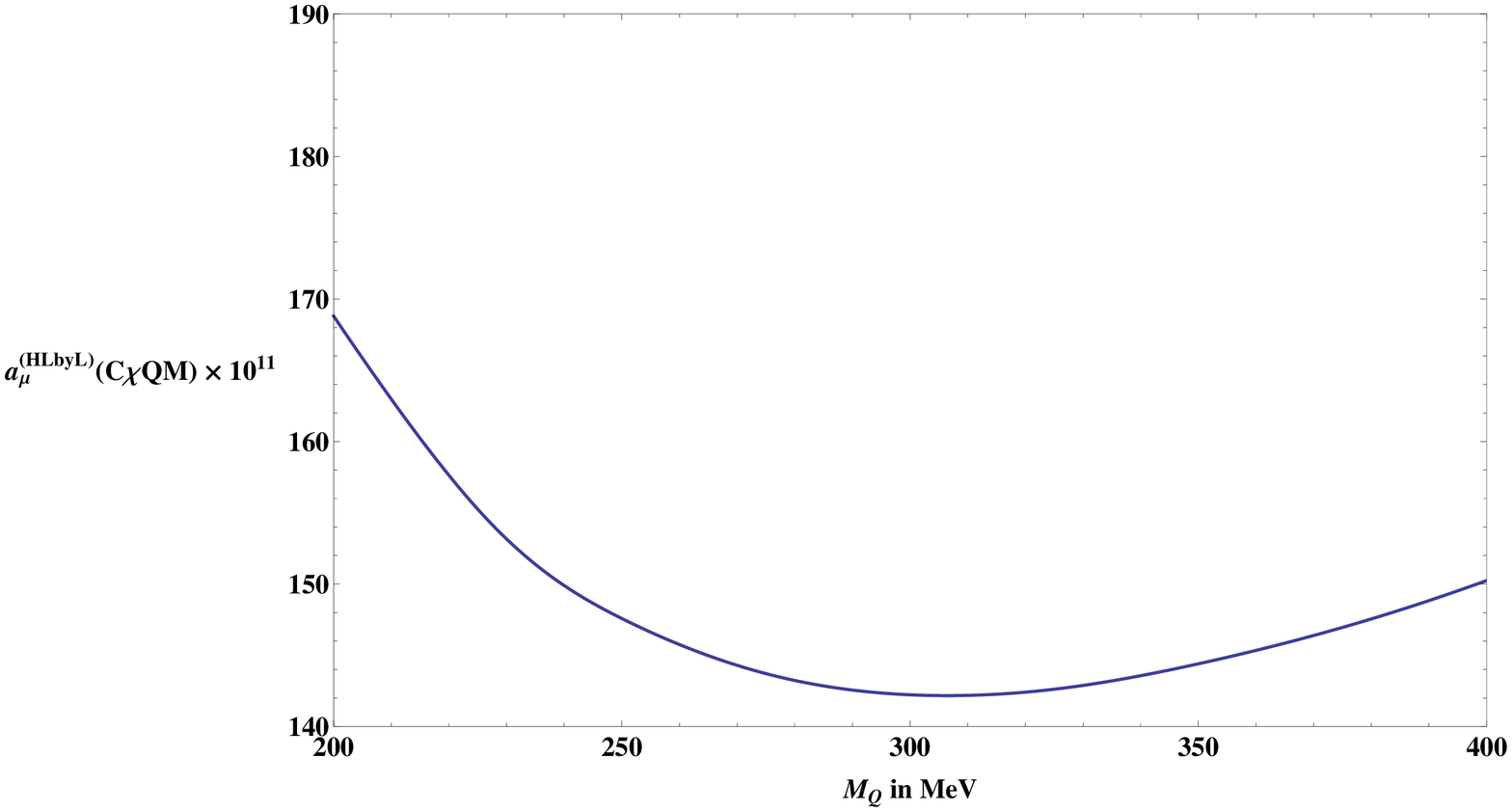}
\bf\caption{\lbl{fig:LbyLtotal}}
\vspace*{0.25cm}
{\it  $a^{(\rm HLbyL)}_{\mu} ({\rm C} \chi {\rm QM})$ versus $M_Q$ in in $10^{-11}$ units.}
\end{center}
\end{figure}
We conclude that a range of values such as
\be
170\le a^{(\rm HLbyL)}_{\mu}\times 10^{11} \le 308\,,
\ee
allowed by the result quoted in Eq.~\rf{germans}, cannot be digested within the C$\chi$QM and in our opinion this casts serious doubts about the compatibility of the model used in ref.~\cite{GFW11} with basic QCD features.
 
\section{\normalsize Hadronic Electroweak  Contributions.}
\setcounter{equation}{0}
\def\theequation{\arabic{section}.\arabic{equation}}
\noi
These are the contributions to the muon anomaly which appear at the two--loop level in the electroweak sector. They are the ones generated by the hadronic
$\gamma\gamma Z$ vertex, with one $\gamma$ and the
$Z$--boson attached to a muon line, as illustrated by the Feynman
diagrams in Fig.~\rf{fig:ggZ}.

\begin{figure}[h]
\begin{center}
\includegraphics[width=0.6\textwidth]{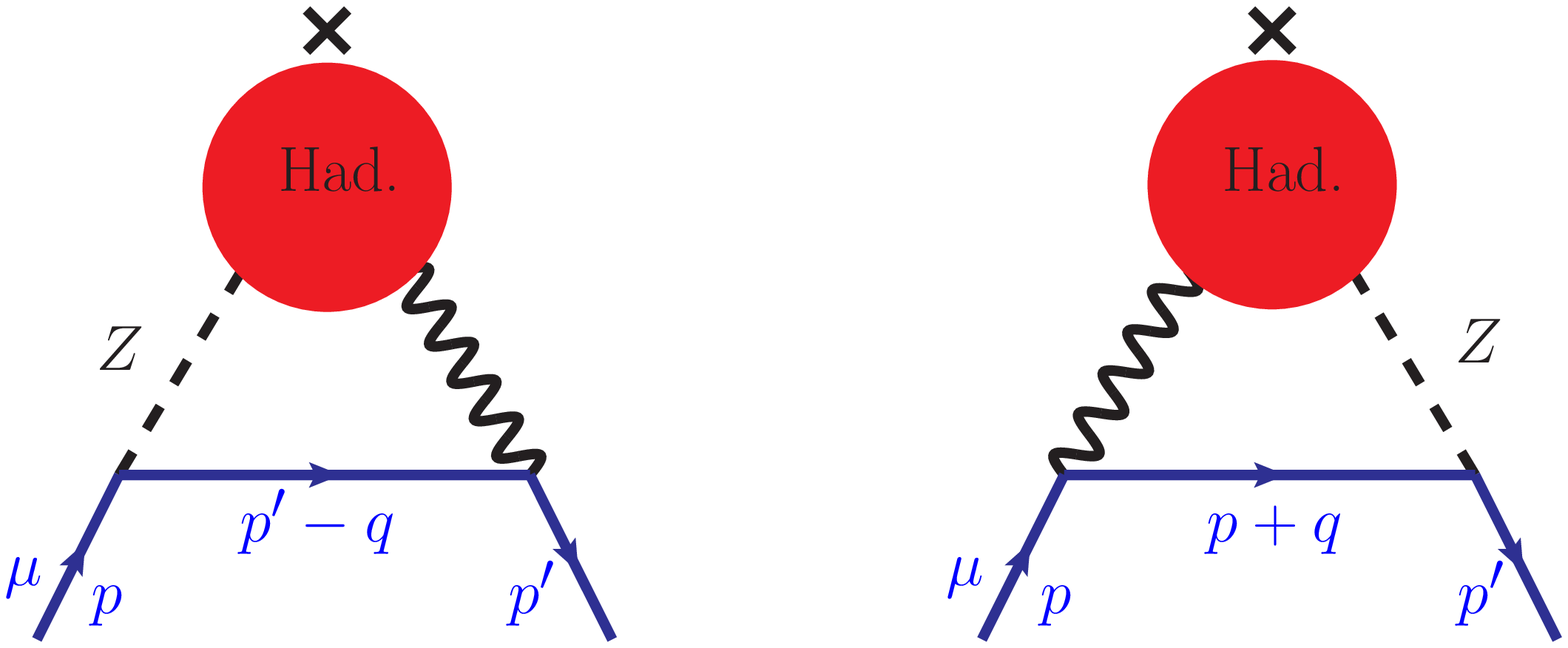}
\bf\caption{\lbl{fig:ggZ}}
\vspace*{0.25cm}
{\it  Feynman diagrams with the hadronic
$\gamma\gamma Z$ vertex which contributes to the muon anomaly.}
\end{center}
\end{figure}

These contributions are
particularly interesting because, {\it a priori}, they could be
enhanced by a large $\log(M_Z^2/m^2_{\mu})$ factor. However, due to the anomaly--free coupling
assignments in the Standard Model,  there is
an important cancellation of UV--scales between the lepton and the quark contributions within a
given family~\cite{PPdeR95,CKM95}. What is left out of this
cancellation in the sector of the $u,d$ and $s$ quarks,  where
the strong interactions play a subtle role at long distances, is governed by the dynamics of spontaneous chiral symmetry breaking~\cite{PPdeR95,KPPdeR02,V03,CMV03,KPPdeR04}. The C$\chi$QM, where the hadronic blob in Fig.~\rf{fig:ggZ} is replaced by a constituent quark loop as illustrated in Fig.~\rf{fig:EWcqm},  offers a simple way to estimate these contributions which we next discuss.

In full generality, the hadronic $\gamma\gamma Z$ contribution to the muon anomly, which we denote by $a_{\mu}^{\rm HEW}$ is given by the following representation~\cite{KPPdeR02}:

{\setl
\bea
a_{\mu}^{\rm (HEW)} & = & (i e^2)\frac{g^2}{16\cos^2\theta_{W}}
\frac{1}{M_{Z}^2}
\lim_{k^2\ra
0}\int\frac{d^4q}{(2\pi)^4}\frac{1}{q^2}
\left(\frac{M_{Z}^2}{q^2 -M_{Z}^2}\right)\times \nn \\
& & 
\frac{1}{4k^2}\tr\left\{(\pslsh+m_{\mu})
\left[\gamma^{\rho}\ksls-\left(k^{\rho}+
\frac{p^{\rho}}{m_{\mu}}\ksls\right)\right]\right. \times \nn \\
 & & 
\left. \left[\gamma^{\mu}\frac{(\pslsh\ -\qsls+m_{\mu})}
{q^2-2q\dd p}
\gamma^{\nu}\gamma_{5}+\gamma^{\nu}\gamma_{5}
\frac{(\pslsh\ +\qsls+m_{\mu})}{q^2+2q\dd p}\gamma^{\mu}
\right]
\right\}W_{\mu\nu\rho}(q,k)\,,
\eea}
 
\noi
where $W_{\mu\nu\rho}(q,k)$ denotes the hadronic Green's function:
\be\lbl{VAV} 
W_{\mu\nu\rho}(q,k)= \int
d^4x\,e^{iq\cdot x}\int d^4y\, e^{i(k-q)\cdot y} \langle
 0\,\vert
T\{V_{\mu}^{\elm}(x)A_{\nu}^{\nc}(y)V_{\rho}^{\elm}(0)\}\vert
0\rangle\,, \ee 
with $k$ the incoming photon four--momentum
associated with the classical external magnetic field, and where
\be
V_{\mu}^{\elm}(x)=\bar{q}(x)\gamma_{\mu}\, Q\, q(x)\,,\qquad\annd\qquad
A_{\nu}^{\nc}(y)=\bar{q}(y)\gamma_{\nu}\gamma_{5} Q_{L}^{(3)}\,,
q(y)
\ee
with 
\be
 Q=Q_{L}=Q_{R}=
{\mbox{\rm \small diag}}\ (2/3,-1/3,-1/3)\,,\qquad\annd\qquad
Q_{L}^{(3)}={\mbox{\rm \small diag}}\
(1,-1,-1)\,.
\ee
The relevant question here is  the contribution to $a_{\mu}^{\rm (HEW)}$ from the
non--anomalous part of $W_{\mu\nu\rho}(q,k)$, denoted by
$\tilde{W}_{\mu\nu\rho}(q,k)$, i.e.
\be\lbl{tilde}
W_{\mu\nu\rho}(q,k)=-
i\frac{N_c}{12\pi^2}\frac{4}{3}\frac{(q-k)_{\nu}}{(q-k)^2}
\epsilon_{\mu\rho\alpha\beta}q^{\alpha}k^{\beta}+\tilde{W}_{\mu\nu\rho}(q,k) \,, \ee 
where the first term in the r.h.s. is the one generated by the VVA anomaly. The second term $\tilde{W}_{\mu\nu\rho}(q,k)$, in the chiral limit where the light quark masses are neglected, is then fully transverse in the axial neutral current ($\nu$ index) and the Ward identities constrain it to have the form $(Q^2 =-q^2)$~\cite{KPPdeR02}:
\be\lbl{mastertilde}
\tilde{W}_{\mu\nu\rho}(q,k)=ik^{\sigma}\left[q_{\rho}
\epsilon_{\mu\nu\alpha\sigma}q^{\alpha}-q_{\sigma}
\epsilon_{\mu\nu\alpha\rho}q^{\alpha}\right]W(Q^2)\,, 
\ee
with only one invariant function $W(Q^2)$ which depends on the details of the dynamics.

\begin{figure}[h]
\begin{center}
\includegraphics[width=0.3\textwidth]{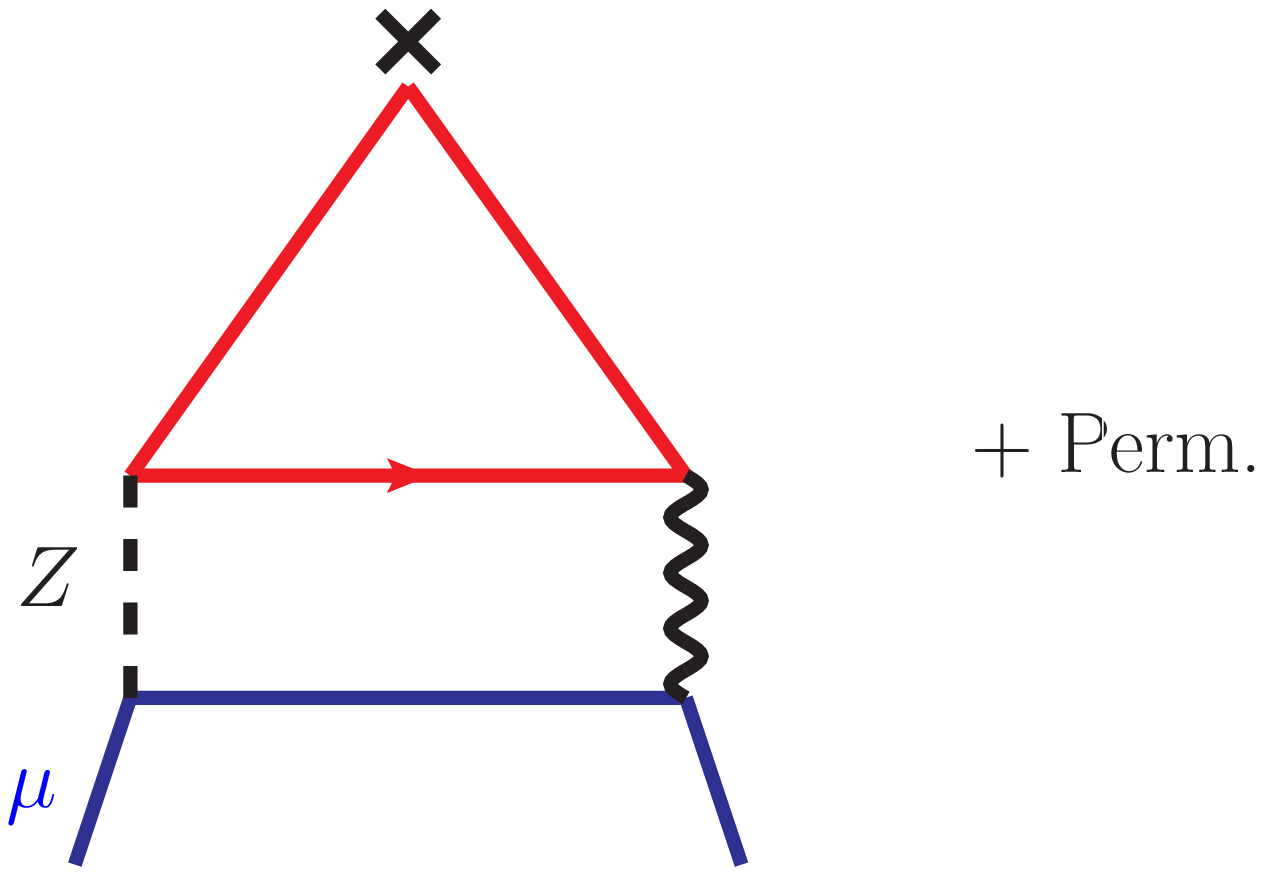}
\bf\caption{\lbl{fig:EWcqm}}
\vspace*{0.25cm}
{\it\small   Feynman diagrams in the C$\chi$QM of the
$\gamma\gamma Z$ vertex type.}
\end{center}
\end{figure}

In the C$\chi$QM, with $g_A =1$, the function $W(Q^2)$ is given by the expression:

{\setl 
\bea\lbl{WchiQM} 
W_{\chi{\rm QM}}(Q^2) &
= & \frac{N_c}{12\pi^2}\frac{8}{3} \frac{1}{M_Q^2}
\times \int_{0}^{1}dx \int_{0}^{1-x}dy
\frac{xy-y(1-y)}{1+\frac{Q^2}{M_{Q}^2} y(1-y)} \nn \\
 & = & \lbl{qmw} \frac{-N_c}{12\pi^2}\frac{2}{3}
\frac{1}{Q^2}\left\{ 1+\frac{M_{Q}^2}{Q^2}
\frac{2}{\sqrt{1+\frac{4M_{Q}^2}{Q^2}}}
\log{\frac{\sqrt{1+\frac{4M_{Q}^2}{Q^2}}-1}
{\sqrt{1+\frac{4M_{Q}^2}{Q^2}}+1}}\right\}\,. 
\eea}

\noi
Not surprisingly, the second term in the brackets coincides with the analytic expression of the C$\chi$QM vertex function given by the term in brackets in the second line of Eq.~\rf{QQvertex}. This is because, up to an overall factor, $W_{\chi{\rm QM}}(Q^2)$
is precisely the same function as $\cF_{{\pi^0}^* \gamma^* \gamma^*}^{(\chi{\rm QM})}\left(q^2,q^2 ,0 \right)$ with the anomaly $(M_Q \ra \infty)$ subtracted. It is this fact that guarantees that $W_{\chi{\rm QM}}(Q^2)$, for $g_A =1$, has  the correct pQCD leading short--distance behaviour~\cite{V03,KPPdeR04}:
\be
W_{\chi{\rm QM}}(Q^2)\ \xrightarrow[Q^2 \ra\infty]{}\ \frac{-N_c}{12\pi^2}\frac{2}{3}
\frac{1}{Q^2}\,.
\ee
The C$\chi$QM, however, has its limitations and does not predict correctly the subleading short--distance behaviour:
\be
\lim_{Q^2 \ra \infty} W_{\chi{\rm QM}}(Q^2)= \frac{-N_c}{12\pi^2}\frac{2}{3}
\left[\frac{1}{Q^2}-2\frac{M_Q^2}{Q^4}\log\frac{Q^2}{M_Q^2}+\cO\left(\frac{M_Q^4}{Q^6}\log\frac{Q^2}{M_Q^2} \right)\right]\,,
\ee
which falls as $\cO\left(\frac{1}{Q^4}\right)$, while the OPE in QCD predicts that this subleading term must fall as  $\cO\left(\frac{1}{Q^6}\right)$~\cite{KPPdeR02}.

Another interesting limit is the long--distance behaviour of the function $W(Q^2)$ which in QCD is related to a coupling constant of $\cO(p^6)$ in the odd--parity sector of the effective chiral Lagrangian~\cite{KPPdeR04}. Here, the prediction of the C$\chi$QM is
\be
W_{\chi{\rm QM}}(Q^2)\ \xrightarrow[Q^2 \ra0]{}\ \frac{-N_c}{12\pi^2}
\frac{1}{9 M_Q^2}\,.
\ee
Unfortunately, there is no model independent prediction for $W(0)$ to compare with.

The contribution to $a_{\mu}^{\rm (HEW)}$ from the 	anomalous term in Eq.~\rf{tilde}, evaluated in the Feynman gauge~\cite{PPdeR95,KPPdeR02} is:

{\setl
\bea
a_{\mu}^{\rm (HEW)}\Big\vert_{\rm\footnotesize anom} & = & 
\frac{\GF}{\sqrt{2}}
\frac{m_{\mu}^2}{8\pi^2}\frac{\alpha}{\pi}\frac{N_c}{3}
\left\{\frac{4}{3}\log\frac{M_Z^2}{m_{\mu}^2}+\frac{2}{3}+\cO\left(\frac{m_{\mu}^2}{M_Z^2}\log\frac{M_Z^2}{m_{\mu}^2} \right) \right\}\nn \\
& = & \frac{m_{\mu}^2}{8\pi^2}\frac{\alpha}{\pi}\times 18.69\,,
\eea}

\noi
and the one from the transverse component $\tilde{W}_{\mu\nu\rho}(q,k)$ in Eq.~\rf{tilde}, evaluated in the C$\chi$QM in the Feynman gauge and with $g_A =1$:
\be\lbl{CchiQM}
a_{\mu}^{\rm (HEW)}\Big\vert_{\mbox{\rm\tiny
transv}}^{\chi{\rm QM}}  = \frac{\GF}{\sqrt{2}}\
\frac{m_{\mu}^2}{8\pi^2}\ \frac{\alpha}{\pi}
\times \frac{N_c}{3}\left[\frac{2}{3}
\log\frac{M_{Z}^2}{M_{Q}^2}-\frac{4}{3}+
\cO\left(\frac{M_{Q}^2}{M_{Z}^2}\log\frac{M_{Z}^2}{M_{Q}^2}\right)\,.
\right]\,.
\ee
The sum of these two contributions takes care of the hadronic sector induced by the dynamics of the light quarks $u$, $d$ and $s$; but, as already mentioned, it is only when added to the lepton contributions from the electron and the muon and the one from the heavy charm quark that the whole sum is gauge independent and it then makes sense in the Standard Model. We reproduce below the details of this overall result:

{\setl
\bea\lbl{overall}
& &  a_{\mu}^{\rm (HEW)}\Big\vert_{e,u,d;\mu,s,c}  =  \frac{\GF}{\sqrt{2}}\
\frac{m_{\mu}^2}{8\pi^2}\ \frac{\alpha}{\pi}\times \nn\\
 & & \left\{ \underbrace{-3 \log\frac{M_Z^2}{m_{\mu}^2}-\frac{5}{2}}_{\footnotesize{\rm electron}}+
 \underbrace{2\log\frac{M_Z^2}{m_{\mu}^2}+1}_{\footnotesize{\rm u,d~~anom}}
 +  \underbrace{\log\frac{M_Z^2}{M_Q^2}-2 }_{\footnotesize{\rm u,d~~transv}~{\rm C}\chi{\rm QM}} \right. \nn \\
 &  & \left.  \underbrace{-3 \log\frac{M_Z^2}{m_{\mu}^2}-\frac{11}{6}+\frac{8}{9}\pi^2}_{\footnotesize{\rm muon}}\ \ 
 \underbrace{-\frac{2}{3}\log\frac{M_Z^2}{m_{\mu}^2}-\frac{1}{3}}_{\footnotesize{\rm s~~anom}}\ \   \underbrace{-\frac{1}{3}\log\frac{M_Z^2}{M_Q^2}+\frac{2}{3} }_{\footnotesize{\rm s~~transv}~{\rm C}\chi{\rm QM}}\ \ 
 \underbrace{+\ 4 \log\frac{M_Z^2}{m_c^2} }_{\footnotesize{\rm charm }} \right\} \,.
\eea}

\noi
Notice how the $\log M_Z^2$ dependence cancels in each generation~\cite{CMV03}, and we finally obtain

{\setl
\bea
a_{\mu}^{\rm (HEW)}\Big\vert_{e,u,d;\mu,s,c}({\rm C}\chi{\rm{QM}}) & = &  \frac{\GF}{\sqrt{2}}\
\frac{m_{\mu}^2}{8\pi^2}\ \frac{\alpha}{\pi}
\left(-\frac{2}{3}\log\frac{M_Q^2}{m_{\mu}^2}-4\log\frac{m_c^2}{m_{\mu}^2} -5+\frac{8}{9}\pi^2 \right)\\
 & = & -\frac{\GF}{\sqrt{2}}\
\frac{m_{\mu}^2}{8\pi^2}\ \frac{\alpha}{\pi}\times\ 18.6 = -5.0\times 10^{-11}\,, 
\eea}

\noi
for $m_c =1.5~\GeV$ and $M_Q =240~\MeV$, a result which within the systematic errors of the model, is compatible with the phenomenological determination~\cite{CMV03}:
\be
a_{\mu}^{\rm (HEW)}\Big\vert_{e,u,d;\mu,s,c}=(-6.7\pm 0.5)\times 10^{-11}\,.
\ee

\section{\normalsize Summary and Conclusions.}
\setcounter{equation}{0}
\def\theequation{\arabic{section}.\arabic{equation}}
From the previous considerations we conclude that the C$\chi$QM provides a useful and simple {\it reference model} to evaluate the hadronic contributions to the anomalous magnetic moment of the muon.  The effective Lagrangian of this model is renormalizable  in the Large--${\rm N_c}$ limit~\cite{Wei10} and, as shown in~\cite{EdeR11},  the number of the required counterterms in this limit is minimized for a choice of the axial coupling: $g_A =1$. 
The only free parameter of the model is then the mass of the constituent quark mass $M_Q$ which in Section II, from a comparison with the phenomenological determination of the lowest order hadronic vacuum polarization contribution to the muon anomaly, has been fixed to 
\be
M_Q =(240\pm 10)~\MeV\,,
\ee
This range of values for $M_Q$ reproduces the phenomenological determination within an error of less than $10\%$. All the other hadronic contributions have then been evaluated for this range of values of $M_Q$ with the results which are summarized in Table~2. 

\begin{table*}[h]
\caption[Results]{{\it Summary of results  for the hadronic contributions to the muon anomly in the C$\chi$QM.}}
\lbl{table1}
\begin{center}
\begin{tabular}{|c|c|} \hline \hline  {\bf Class} &
{\bf Result in $10^{-10}$ units}
\\ \hline \hline
HVP & $ 652\begin{array}{r} +47 \\ -42  \end{array}$
\\  HVP to $\cO\left(\frac{\alpha}{\pi}\right)^3$  & $ -6.4 \pm 1.2$\\
 HLbyL  & $ 15.0 \pm 0.3$ \\
HEW & $ -0.5$  \\
\hline\hline
\end{tabular}
\end{center}
\end{table*} 

\noi
We want to emphasize that the errors quoted in Table~2 are only those generated by the error of $M_Q$ in Eq.~\rf{MQ} and they do not reflect the systematic error of the model. These results, within a systematic error of 20\% to 30\%, are in good agreement with the phenomenological determinations. One exception, discussed in detail in Section  III.4.1,   is the contribution from the $\pi^0 \gamma$ intermediate state to hadronic vacuum polarization where the C$\chi$QM, because of its $SU(3)$ invariance, fails to reproduce the phenomenological determination which is particularly enhanced because of the large observed branching ratio in Eq.~\rf{omega}.

Ironically, the error $\pm 0.3$ for the Hadronic Light--by--Light contribution appears to be the smallest relative error. This is due to the fact that, as shown in Fig.~16,  the sum of the quark loop contribution and the Goldstone exchange contribution for values of $M_Q$ in the range of Eq.~\rf{MQ} is already very near to the minimum in the $M_Q$--dependence of the sum of these two contributions, which occurs at $M_Q \simeq 300~\MeV$. In other words, in the C$\chi$QM the contribution to the muon anomaly which is less sensitive to the value of the constituent quark mass is precisely the one from the hadronic light--by--light scattering. This fact, however, should not mask the intrinsic systematic error which has not been included. Within an expected systematic error of $\sim 20$\%, our results agree with the phenomenological determinations reviewed in ref.~\cite{PdeRV10}. An exception, however,  is the determination quoted in Eq.~\rf{germans}. As discussed in the text, the large range of values allowed by this result, cannot be digested within the C$\chi$QM and in our opinion casts serious doubts about the compatibility of the model used in ref.~\cite{GFW11} with basic QCD features.

Another interesting feature, which has appeared when evaluating the hadronic electroweak contributions, is the impact of the choice $g_A =1$, which was initially made on theoretical grounds. It turns out that it is only for this choice that the C$\chi$QM has the correct matching at short--distances with the one predicted by the OPE in QCD~\cite{V03,KPPdeR04} when evaluating the hadronic electroweak contribution.

Concerning the next--to--leading contributions from Hadronic Vacuum Polarization we have made two observations which are in fact model independent. On the one hand we have explained why this contribution is smaller than the naive expected order of magnitude and on the other hand we have derived a sum rule in Eq.~\rf{dyncon} which offers an interesting constraint when evaluating radiative corrections. 
  

\newpage 

\acknowledgments

We wish to thank Marc Knecht for many helpful discussions on the topics discussed in this paper. We thank Marc Knecht, Santi Peris and Laurent Lellouch for a careful reading of the manuscript

The work of DG has been supported by MICINN (grant FPA2009-09638) and DGIID-DGA
(grant 2009-E24/2) and by the Spanish Consolider-Ingenio 2010 Program
CPAN (CSD2007-00042).


\vfill

\end{document}